\documentclass[pdflatex,sn-nature]{sn-jnl}
\usepackage{graphicx}
\usepackage{amsmath,amssymb,amsfonts}
\usepackage{xcolor}
\usepackage{booktabs}
\usepackage{longtable,rotating,booktabs}
\usepackage{pdflscape}
\usepackage{aas_macros}

\begin{document}

%% ============================================================
%% TITLE
%% ============================================================

% \title{There is no single density: star-forming regions and galaxies hold more dense gas than long assumed}
\title{There is no single density: star-forming regions and galaxies hold more dense ionized gas than long assumed}
%% ---------------------- Autores ----------------------------
\author*[1]{\fnm{J. Eduardo} \sur{Méndez-Delgado}}\email{jmendez@astro.unam.mx}
\author[2]{\fnm{Christophe} \sur{Morisset}}
\author[3]{\fnm{William J.} \sur{Henney}}
\author[4]{\fnm{Niv} \sur{Drory}}
\author[5]{\fnm{Oleg V.} \sur{Egorov}}
\author[1,6]{\fnm{Sebastián F.} \sur{Sánchez}}
\author[7,8]{\fnm{Juna A.} \sur{Kollmeier}}
\author[5]{\fnm{Kathryn} \sur{Kreckel}}
\author[7,9]{\fnm{Guillermo} \sur{Blanc}}
\author[10]{\fnm{Grażyna} \sur{Stasińska}}
\author[11]{\fnm{Evelyn J.} \sur{Johnston}}
\author[1]{\fnm{Héctor J.} \sur{Ibarra-Medel}}
\author[12]{\fnm{Alfredo J.} \sur{Mejía-Narváez}}
\author[13,6]{\fnm{César} \sur{Esteban}}
\author[13,6]{\fnm{Jorge} \sur{García-Rojas}}
\author[9]{\fnm{Amrita} \sur{Singh}}
\author[14,15,16]{\fnm{Ivan Yu.} \sur{Katkov}}
\author[17]{\fnm{Evan D.} \sur{Skillman}}
\author[3]{\fnm{Rogelio} \sur{Orozco-Duarte}}
\author[5,18]{\fnm{Igor A.} \sur{Zinchenko}}
\author[19,20]{\fnm{Alejandra Z.} \sur{Lugo-Aranda}}
\author[19]{\fnm{Aida} \sur{Wofford}}
\author[21]{\fnm{Simon C. O.} \sur{Glover}}
\author[5]{\fnm{Evgeniya} \sur{Egorova}}
\author[1]{\fnm{Rodolfo de J.} \sur{Zermeño}}
\author[1]{\fnm{Lesly C.} \sur{Castañeda-Carlos}}
\author[5]{\fnm{Fu-Heng} \sur{Liang}}
\author[5]{\fnm{Natascha} \sur{Sattler}}
\author[1]{\fnm{Irene} \sur{Cruz-González}}
\author[22]{\fnm{Joel R.} \sur{Brownstein}}
\author[23]{\fnm{Thomas} \sur{Hilder}}
\author[24,25]{\fnm{Donald P.} \sur{Schneider}}
 
%% -------------------- Afiliaciones -------------------------
\affil[1]{Instituto de Astronomía, Universidad Nacional Autónoma de México, A.P. 70-264, 04510 Ciudad de México, México}
\affil[2]{Instituto de Ciencias Físicas, Universidad Nacional Autónoma de México, Av. Universidad s/n, 62210 Cuernavaca, Morelos, México}
\affil[3]{Instituto de Radioastronomía y Astrofísica, Universidad Nacional Autónoma de México, Antigua Carretera a Pátzcuaro 8701, 58089 Morelia, Michoacán, México}
\affil[4]{McDonald Observatory, The University of Texas at Austin, 1 University Station, Austin, TX 78712-0259, USA}
\affil[5]{Astronomisches Rechen-Institut, Zentrum für Astronomie der Universität Heidelberg, Mönchhofstr. 12-14, 69120 Heidelberg, Germany}
\affil[6]{Instituto de Astrofísica de Canarias, Vía Láctea s/n, 38205 La Laguna, Tenerife, Spain}
\affil[7]{Observatories of the Carnegie Institution for Science, 813 Santa Barbara Street, Pasadena, CA 91101, USA}
\affil[8]{Canadian Institute for Theoretical Astrophysics (CITA), University of Toronto, 60 St George St, Toronto, ON M5S 3H8, Canada}
\affil[9]{Departamento de Astronomía, Universidad de Chile, Camino del Observatorio 1515, Las Condes, Santiago, Chile}
\affil[10]{LUX, Observatoire de Paris, Université PSL, Sorbonne Université, CNRS, 92190 Meudon, France}
\affil[11]{Instituto de Estudios Astrofísicos, Facultad de Ingeniería y Ciencias, Universidad Diego Portales, Av. Ejército Libertador 441, Santiago, Chile}
\affil[12]{Universidad de Chile, Av. Libertador Bernardo O'Higgins 1058, Santiago, Chile}
\affil[13]{Departamento de Astrofísica, Universidad de La Laguna, 38206 La Laguna, Tenerife, Spain}
\affil[14]{New York University Abu Dhabi, PO Box 129188, Abu Dhabi, UAE}
\affil[15]{Center for Astrophysics and Space Science (CASS), New York University Abu Dhabi, PO Box 129188, Abu Dhabi, UAE}
\affil[16]{Sternberg Astronomical Institute, Lomonosov Moscow State University, Universitetskij pr. 13, 119234 Moscow, Russia}
\affil[17]{Minnesota Institute for Astrophysics, University of Minnesota, 116 Church St. SE, Minneapolis, MN 55455, USA}
\affil[18]{Main Astronomical Observatory, National Academy of Sciences of Ukraine, 27 Akademika Zabolotnoho St., 03143 Kyiv, Ukraine}
\affil[19]{Instituto de Astronomía, Universidad Nacional Autónoma de México, A.P. 106, 22800 Ensenada, B.C., México}
\affil[20]{Institute of Astrophysics, Facultad de Ciencias Exactas, Universidad Andrés Bello, Sede Concepción, Talcahuano, Chile}
\affil[21]{Universität Heidelberg, Zentrum für Astronomie, Institut für Theoretische Astrophysik, Albert-Ueberle-Straße 2, 69120 Heidelberg, Germany}
\affil[22]{Department of Physics and Astronomy, University of Utah, 270 S. 1400 E. \#E2108, Salt Lake City, UT 84112, USA}
\affil[23]{School of Physics and Astronomy, Monash University, VIC 3800, Australia}
\affil[24]{Department of Astronomy \& Astrophysics, The Pennsylvania State University, University Park, PA 16802, USA}
\affil[25]{Institute for Gravitation and the Cosmos, The Pennsylvania State University, University Park, PA 16802, USA}

\abstract{
Ionized gas fills star-forming regions and galaxies, and nearly everything we know about its temperature, pressure, mass, and composition is inferred from its emission lines~\cite{Seaton:57,Peimbert:71,Osterbrock:06}. The electron density is needed for all of these, yet a longstanding puzzle has resisted explanation: different density-sensitive lines, applied to the same gas, return values that disagree by up to two orders of magnitude. This is usually attributed either to each line tracing a physically distinct ionization zone or to imperfect atomic data~\cite{JuandeDios:21,Berg:21,Mingozzi:22,Martinez:25}. Here we show that the disagreement is neither a flaw in the atomic data nor an ionization-stratification effect, but something more fundamental. Each diagnostic is tuned to a particular density, and when a nebula contains gas across a wide range of densities — as real nebulae do — each line reports the part of that range it is most sensitive to. The diagnostics do not measure a representative average density; they respond to different parts of a broad density distribution. This resolves the discrepancy with a simple relation between the density each line returns and the density it is most sensitive to — a relation that holds from individual H\,\textsc{ii} regions to whole galaxies, near and far, and reveals that ionized nebulae contain far more dense gas than any one diagnostic implies. A nebula has no single electron density to measure, but a broad density distribution, and the masses, pressures, abundances and energetics built on the single-density assumption must be reconsidered, from nearby star-forming regions to galaxies across cosmic time.

}

\keywords{H\,\textsc{ii} regions, ISM: structure, galaxies: ISM, nebulae: general}

\maketitle

%% ============================================================
%% MAIN TEXT
%% ============================================================

The electron density $n_e$ regulates the cooling, pressure, energetics and emission of ionized astrophysical plasmas. Forbidden-line density diagnostics are therefore among the most fundamental tools in nebular spectroscopy, yet their systematic disagreement has never been satisfactorily explained~\cite{Seaton:57,Peimbert:71,Rubin:89,Copetti:02,Wang:04,Rickards:2024,MendezDelgado:24}. The standard diagnostics [S\,\textsc{ii}]~$\lambda6717/\lambda6731$ and [O\,\textsc{ii}]~$\lambda3727/\lambda3729$ routinely return electron densities of a few hundred cm$^{-3}$ in H\,\textsc{ii} regions and star-forming galaxies, whereas higher-ionization diagnostics such as [Cl\,\textsc{iii}]~$\lambda 5518/\lambda 5538$, [Ar\,\textsc{iv}]~$\lambda4711/\lambda4740$ and the UV intercombination lines C\,\textsc{iii}]~$\lambda1907/\lambda1909$ and Si\,\textsc{iii}]~$\lambda1892/\lambda1882$ return values one to three orders of magnitude larger for the same objects~\cite{Kewley:2019,Steidel:14,Davies:21}. These discrepancies are conventionally attributed to ionization stratification -- the idea that each ion traces a physically distinct gas phase -- \cite{Berg:21,Mingozzi:22,Martinez:25} or to uncertainties in atomic data \cite{Rodriguez:20,JuandeDios:21}, and are typically addressed by adopting a single diagnostic as representative while treating all others as supplementary information. Here we show that this interpretation is fundamentally incomplete: an effect conceptually analogous to the density-dependent line selection long recognized in the emission-line regions of active galaxies~\cite{Filippenko:85,Baldwin:95,Ferguson:97} appears in H\,\textsc{ii} regions and star-forming galaxies, which prove to span a far broader range of densities than has been assumed. The observed hierarchy of inferred densities emerges naturally from the interaction between these broad unresolved density distributions and the intrinsic density-response functions of forbidden-line diagnostics. The latter preferentially sample gas near the density where each ratio is maximally sensitive to changes in $n_e$. Forbidden-line density measurements are therefore intrinsically biased, and the nature of that bias reveals the existence of a dense, emissively important gas component contributing substantially to the emergent forbidden-line luminosity of ionized nebulae.

For each density-sensitive line ratio $R$, we define the maximum-sensitivity density $n_\mathcal{M}$ as the electron density at which the logarithmic derivative of the ratio is maximal:
\begin{equation}
n_\mathcal{M} = {\arg\max}\left|\frac{d\log R}{d\log n_e}\right|.
\label{eq:nM}
\end{equation}
This is a purely atomic quantity, set by the critical densities and transition probabilities of the levels involved. A low-$n_\mathcal{M}$ diagnostic such as [S\,\textsc{ii}]~$\lambda6717/\lambda6731$ ($\log_{10} n_\mathcal{M}\approx3.1$) has its largest logarithmic response near $10^3$~cm$^{-3}$; at substantially lower or higher densities the ratio approaches an asymptotic regime. By contrast, a high-$n_\mathcal{M}$ diagnostic such as C\,\textsc{iii}]~$\lambda1907/\lambda1909$ ($\log_{10} n_\mathcal{M}\approx6.7$) has its response kernel displaced by several orders of magnitude toward higher density. Thus the two ratios do not simply provide noisier or cleaner estimates of the same quantity. In a nebula containing gas over a broad range of densities, the two diagnostics weight different regions of the same unresolved electron-density probability distribution function (PDF). In such a scenario, the density inferred from each ratio is therefore not a direct measurement of a unique physical density, but the result of folding the underlying PDF through a different atomic response function.

\begin{figure*}
\centering
\includegraphics[width=0.92\textwidth]{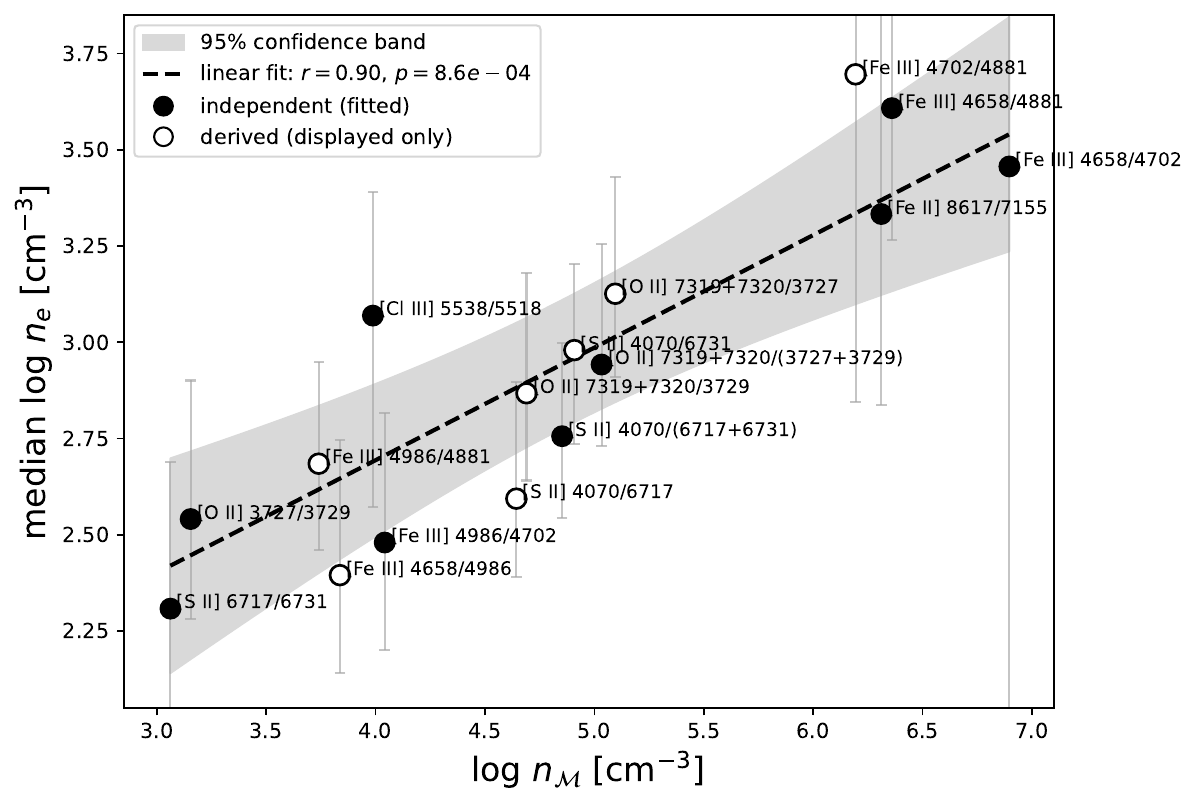}
\caption{
\textbf{Systematic hierarchy of forbidden-line density diagnostics across the Orion Nebula.}
Median electron densities inferred from 16 forbidden-line diagnostics measured spaxel-by-spaxel across the common M42--M43 region observed by SDSS-V/LVM, plotted as a function of the maximum-sensitivity density $n_\mathcal{M}$ of each diagnostic ratio. Diagnostics with progressively larger $n_\mathcal{M}$ systematically recover larger electron densities, defining a tight linear sequence spanning nearly four orders of magnitude in sensitivity density. Filled symbols mark the nine algebraically independent diagnostics used in the fit; open symbols mark the seven that are exact combinations of these and are displayed for completeness only. Grey bars show the 16th--84th percentile range of the spaxel-to-spaxel density distribution for each diagnostic --- the spatial dispersion across the region, not a measurement uncertainty; the statistical uncertainty on each median is negligible relative to the marker size given the 1226 spaxels, and the linear fit is performed on the medians. The dashed line shows the best-fitting linear relation of equation~\eqref{eq:orion_fit} and the shaded region its 95\% confidence band, with Pearson correlation coefficient $r = 0.90$ and $p = 8.6\times10^{-4}$. The existence of a continuous ordered sequence across ions, ionization potentials and wavelength regimes demonstrates that forbidden-line density diagnostics do not measure a unique physical density, but selectively weight different regions of the same unresolved electron-density probability distribution --- revealing the existence of a dense, emissively important gas component that low-$n_\mathcal{M}$ diagnostics are insensitive to by construction.
}
\label{fig:main_sequence}
\end{figure*}

The first and most direct observational evidence for this picture comes from the Orion Nebula, the nearest massive star-forming region and the most thoroughly studied H\,\textsc{ii} region on the sky~\cite{Odell:2001}. New observations from the Local Volume Mapper (LVM; \cite{Drory:2024}) survey of SDSS-V~\cite{Kollmeier:2026} provide spatially resolved spectroscopy across the full M42--M43 region with wavelength coverage $\sim3600$--9800$~\AA$ at a spatial scale of $\sim0.07$~pc~\cite{Kreckel:2024}. The combination of broad simultaneous wavelength coverage and contiguous spatial mapping is decisive: it allows us to derive resolved density maps from 16 forbidden-line density diagnostics within exactly the same physical 1226 spaxels, spanning nearly four orders of magnitude in $n_\mathcal{M}$, with no differential aperture or cross-calibration effects and no ambiguity about whether diagnostics sample the same projected emitting gas.

Fig.~\ref{fig:main_sequence} shows the central result. The median electron density inferred from each diagnostic across the 1226 individual spaxels follows a remarkably linear sequence with $\log_{10} n_\mathcal{M}$, continuous and ordered across transitions spanning widely different wavelengths, ionization potentials and critical densities. This sequence is described by
\begin{equation}
\log\,n_{e,\mathrm{obs}} = (0.29 \pm 0.05)\,\log\,n_\mathcal{M} + (1.52 \pm 0.27),
\label{eq:orion_fit}
\end{equation}
with Pearson correlation coefficient $r = 0.90$ and $p = 8.6\times10^{-4}$, fitted to the nine algebraically independent diagnostics (see Methods). The slope is not zero --- which would imply all diagnostics measure the same density --- and it is not unity --- which would imply each diagnostic simply returns its own $n_\mathcal{M}$ regardless of the gas. The intermediate value of $\approx0.3$ is the direct signature of a broad, unresolved density distribution: each diagnostic recovers a density between the low-density bulk and the high-density tail of the nebular PDF, weighted by its response kernel. The two conventional explanations for diagnostic discrepancies are ruled out here. Multiple distinct ratios of the same ionic species --- four diagnostics of S$^+$, four of O$^+$ and six of Fe$^{2+}$ --- show that the hierarchy persists even when the ionization structure and relative abundances are essentially fixed, ruling out ionization stratification as its primary driver. This does not imply that ionization stratification is absent, but it cannot be the organizing principle behind the observed sequence. Uncertainties in atomic data can introduce systematic offsets between individual diagnostics~\cite{Rodriguez:20,JuandeDios:21}, but cannot plausibly produce the continuous, monotonic and highly correlated sequence spanning different elements, different ionization stages and independent sets of atomic calculations. The observed sequence does not contradict the atomic physics of the diagnostics --- it emerges directly from it. 

\begin{figure*}
\centering
\includegraphics[width=0.99\textwidth]{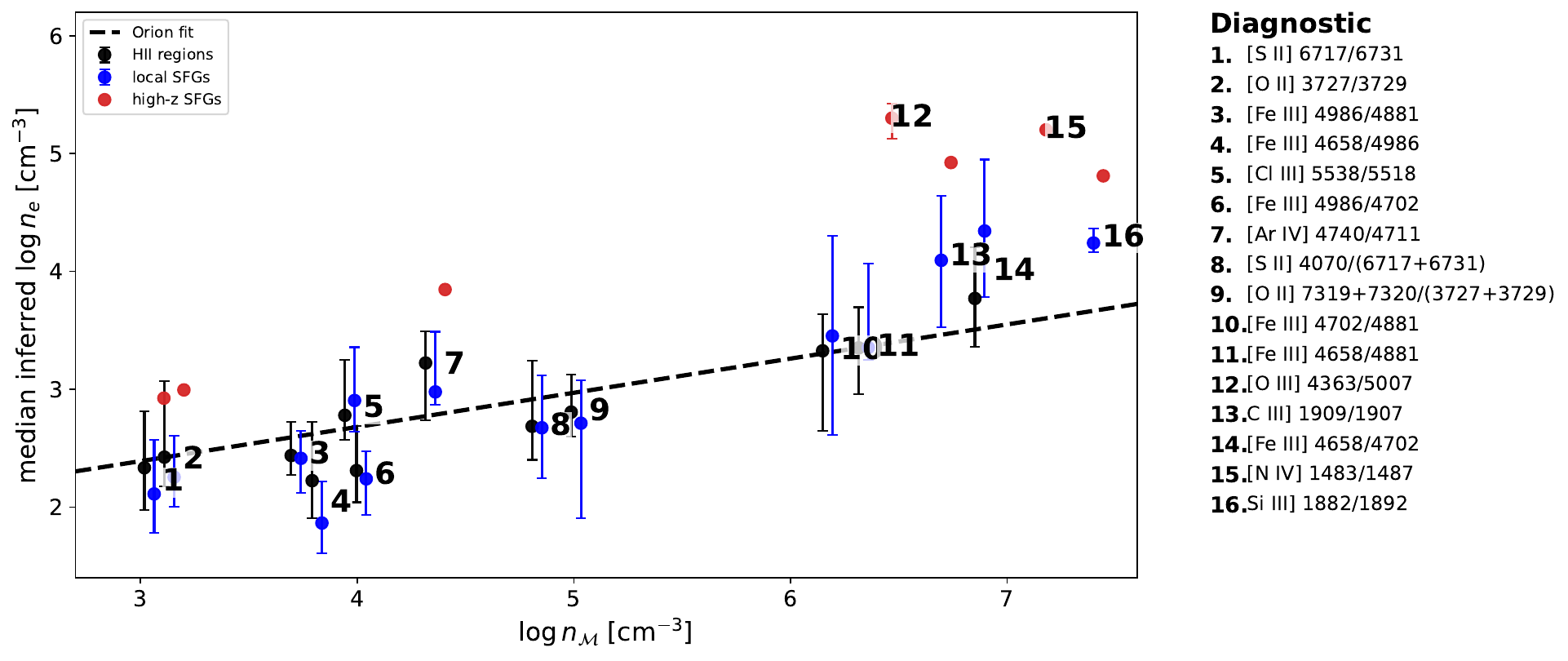}
\caption{
\textbf{Density diagnostics follow the same sensitivity ordering across nebular environments.}
Median inferred electron density as a function of the maximum-sensitivity density $n_\mathcal{M}$ for H\,\textsc{ii} regions (black points) and local star-forming galaxies (blue points) from the DESIRED database~\cite{MendezDelgado:23b}, together with high-redshift star-forming galaxies (red points) from the literature. The dashed line shows the relation measured in the Orion Nebula (equation~\ref{eq:orion_fit}) and is not refitted to the external samples. Error bars show the 16th--84th percentile range of the inferred density distribution across individual objects in each sample. Across local H\,\textsc{ii} regions and nearby galaxies, diagnostics with larger $n_\mathcal{M}$ recover systematically larger median densities, extending the Orion density hierarchy to integrated nebular spectra.  At high redshift, the available sample is limited, but the measurements qualitatively follow the expected hierarchy, with inferred densities falling above the Orion relation in a manner consistent with broader or denser underlying density distributions at earlier cosmic epochs.}
\label{fig:universal}
\end{figure*}

The Orion sequence is not a peculiarity of the nearest H\,\textsc{ii} region. Fig.~\ref{fig:universal} shows that the same ordered hierarchy --- with roughly the same slope --- is recovered across a broad sample of extragalactic H\,\textsc{ii} regions and star-forming galaxies from the DESIRED database~\cite{MendezDelgado:23b}, spanning a wide range of metallicities, ionization conditions and other physical properties. Each nebula has its own characteristic density scale: [S\,\textsc{ii}]~$\lambda6717/\lambda6731$ returns different absolute values across different environments depending on the bulk conditions of the gas. Yet within each environment, the remaining diagnostics return systematically higher densities in the same ordered sequence, with a slope consistent, on average, with that of equation~(\ref{eq:orion_fit}). The offset between diagnostics is not a quirk of Orion's geometry, ionization structure or proximity --- it is a property of the diagnostics themselves, operating wherever ionized gas contains an unresolved density distribution broader than the response kernel of any individual ratio. The same qualitative ordering is also recovered in planetary nebulae, with a shallower slope and higher intercept reflecting their higher characteristic densities and distinct density structure; as that population lies outside the calibration of equation~(\ref{eq:orion_fit}), it is analysed separately (M\'endez-Delgado et al., in preparation).

The relation appears to extend to high-redshift star-forming galaxies, where the accessible diagnostics range from [S\,\textsc{ii}]~$\lambda6717/\lambda6731$ to high-$n_\mathcal{M}$ UV intercombination lines and [O\,\textsc{iii}]~$\lambda4363/\lambda5007$~\cite{Arellano:26}. Although the high-redshift statistics remain limited, the available measurements reproduce the local hierarchy, with median values falling above the Orion relation as expected for broader or denser underlying distributions at earlier cosmic epochs. Larger samples with simultaneous multi-diagnostic coverage, now becoming accessible with JWST, will be essential to establish whether the observed density hierarchy evolves with redshift. Regardless, the density-selection bias documented here seems to remain operative at all epochs: densities inferred from high-$n_\mathcal{M}$ diagnostics overestimate the bulk gas density, and comparisons between low-$z$ and high-$z$ measurements made with different diagnostics remain systematically compromised unless the bias is explicitly accounted for.

\begin{figure}
\centering
\includegraphics[width=0.92\columnwidth]{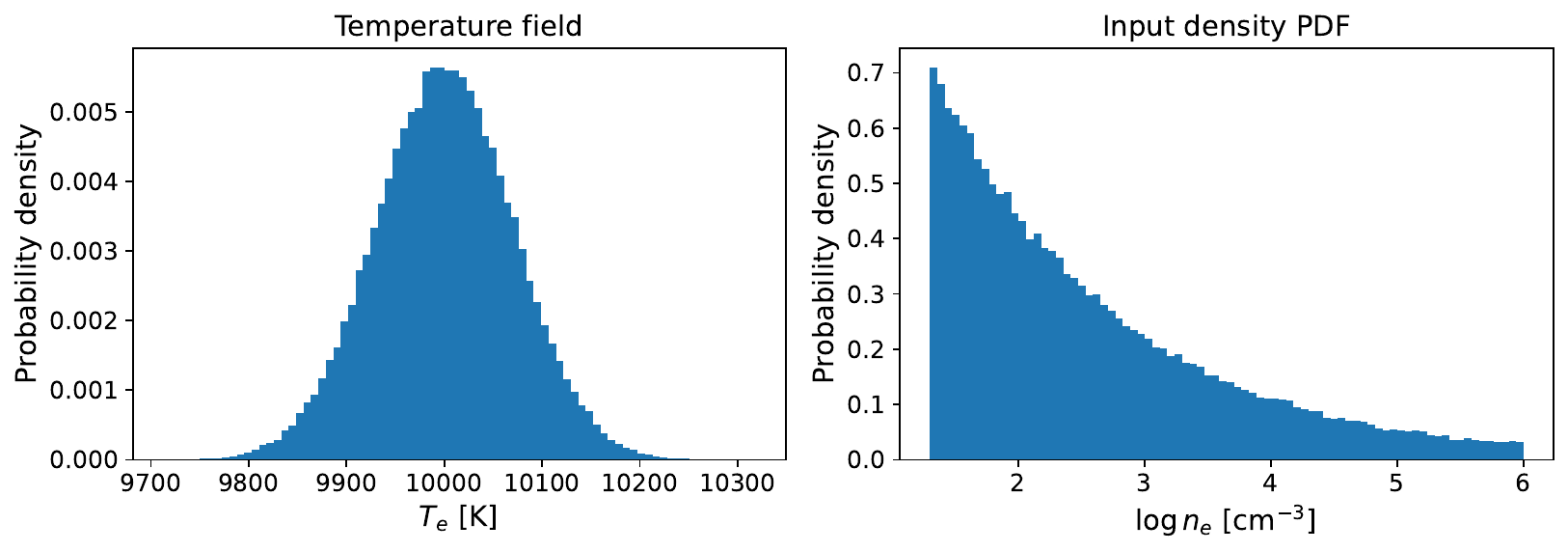}
\caption{
\textbf{Input distributions of the forward density model.}
Left: electron temperature distribution adopted in the simulations, narrowly distributed around $T_e = 10^4$~K in order to isolate the effects of density structure on the inferred diagnostics. Right: input emission-measure density distribution $n_e H_n \propto n_e^{\beta+1}$ with slope $\beta = -1.3$, spanning from $n_\mathrm{min} = 10^{1.3}$~cm$^{-3}$ to $n_\mathrm{max} = 10^{6}$~cm$^{-3}$, where $H_n\,dn \equiv n_e^2\,dV$ is the contribution of gas in $[n_e, n_e+dn]$ to the total emission measure, so that the emission-measure weighting $n_e^2$ is already included in $H_n$. The power law is adopted as the minimal-parameter description of a broad continuous density distribution; the slope $\beta$ is treated as a free parameter. No ionization stratification, phase structure or spatial correlations are imposed.
}
\label{fig:model_input}
\end{figure}

\begin{figure}
\centering
\includegraphics[width=0.92\columnwidth]{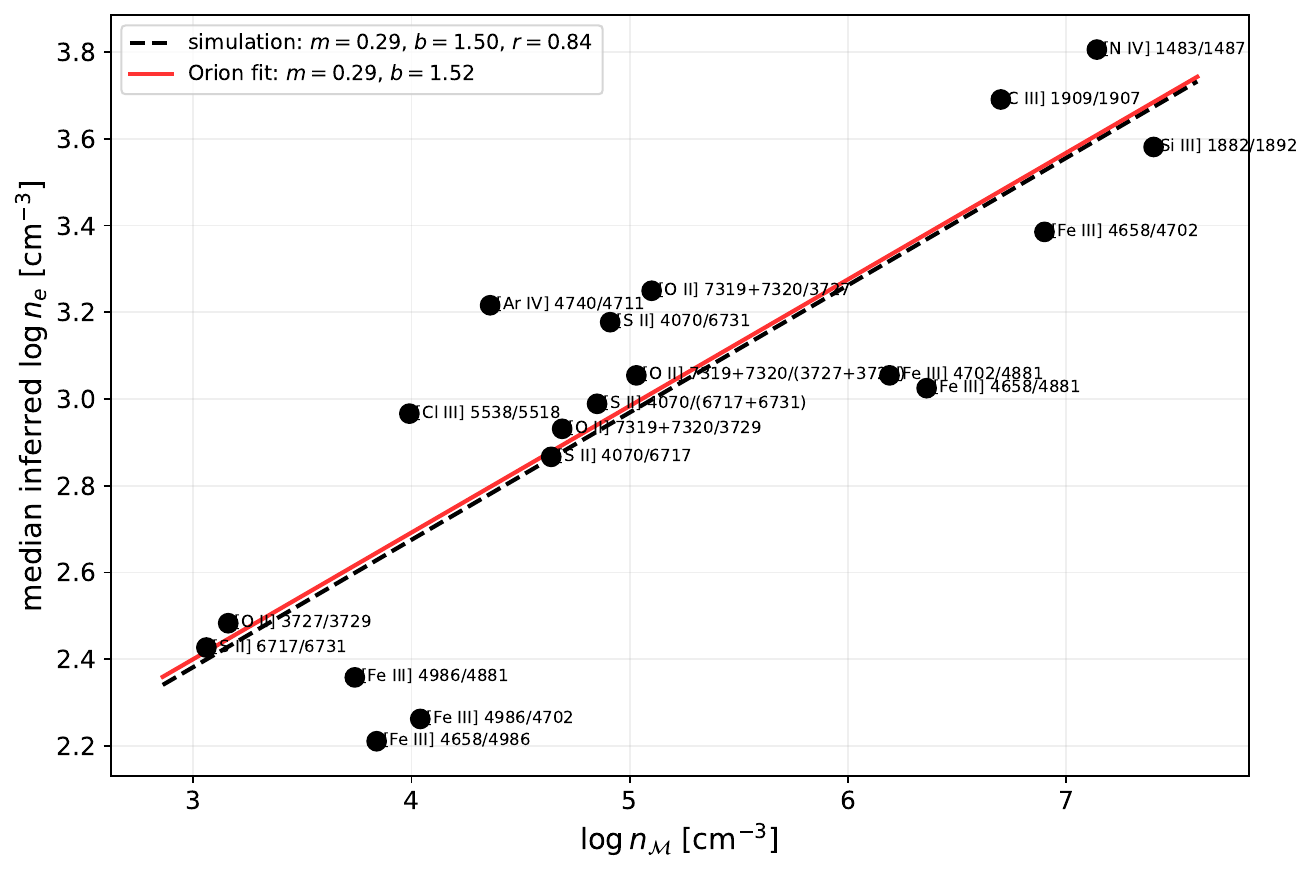}
\caption{
\textbf{Forward model reproduces the observed density hierarchy.}
Median inferred electron density as a function of the maximum-sensitivity density $n_\mathcal{M}$ for the forward model (black dashed line; $m = 0.29$, $b = 1.50$, $r = 0.84$) compared to the observed Orion relation of equation~(\ref{eq:orion_fit}) (red line; $m = 0.29$, $b = 1.52$). The model contains no ionization stratification and no discrete gas phases. The density field follows a power-law emission-measure distribution $H_n \propto n_e^{\beta}$ with $\beta = -1.3$, $n_\mathrm{min} = 10^{1.3}$~cm$^{-3}$ and $n_\mathrm{max} = 10^6$~cm$^{-3}$ (equation~\ref{eq:powerlaw_methods}). The close agreement demonstrates that the observed density hierarchy arises naturally from broad unresolved density distributions viewed through diagnostics with different atomic response functions. Within the context of the power-law emission-measure model, the slope of the relation primarily traces the power-law index $\beta$ and the density range spanned, whereas the intercept reflects the characteristic density scale set by $n_\mathrm{min}$.
}
\label{fig:model_sequence}
\end{figure}

The interpretation of the slope and intercept of equation~(\ref{eq:orion_fit}) depends on the functional form of the underlying density distribution. Within the specific case of power-law emission-measure distributions, the intercept primarily reflects the characteristic density scale set by $n_\mathrm{min}$, while the slope encodes its width: broader distributions produce steeper relations because a larger fraction of the forbidden-line luminosity originates from the dense tail, which high-$n_\mathcal{M}$ diagnostics are preferentially sensitive to. At the limits, a vanishing slope corresponds to a delta-function distribution with no dense tail to pull any diagnostic upward, as confirmed by the narrow-PDF test (Extended Data Fig.~\ref{fig:extdata_nobiases}). This limiting case is realized observationally in the Orion Nebula's photoionized Herbig-Haro objects HH\,202-S and HH\,204, where high-resolution spectroscopy isolates a single dense gas component and the diagnostic hierarchy disappears (see Methods). The observed value of $\approx0.3$ therefore implies that gas at $\log _{10} n_e \gtrsim 4$~cm$^{-3}$ contributes meaningfully to the emergent luminosity, yet that no single diagnostic is entirely decoupled from the bulk of the gas.

To test this interpretation quantitatively, we construct a minimal, dust-free forward model in which the electron density at each point in a three-dimensional volume is drawn from a power-law emission-measure distribution (Fig.~\ref{fig:model_input}), chosen as the minimal-parameter description of a broad, continuous density field rather than as a prediction of any specific physical model, while the electron temperature remains nearly isothermal and the upper density is truncated at $\log_{10} n_e = 6$ for definiteness, a choice to which the recovered relation is insensitive. No ionization stratification, phase structure or spatial correlations are imposed. For each diagnostic, we compute \textsc{PyNeb} emissivities~\cite{Luridiana:15}, integrate the numerator and denominator emissivity cubes along the line of sight, and invert the resulting line ratios using the standard density determinations exactly as in observational analyses. The inferred densities therefore correspond not to the input volumetric PDF itself, but to different emissivity-weighted projections of that distribution filtered through the response function of each diagnostic.

Fig.~\ref{fig:model_sequence} shows that this minimal model reproduces the observed Orion sequence with remarkable fidelity ($m = 0.29$, $b = 1.50$, $r = 0.84$, compared to the observed $m = 0.29$, $b = 1.52$), with no tuning of the slope to the data. The density hierarchy is therefore not an artefact of complex geometry or discrete ionization phases, but the natural consequence of a broad unresolved density distribution seen through diagnostics with different response functions. Within the context of the power-law emission-measure models explored here, the observed Orion relation is reproduced by distributions with slope $\beta \approx -1.3$ spanning nearly five orders of magnitude in density, implying that the ionized gas encompasses a broad range of structures from diffuse interclump gas to dense compact clumps within the same emitting volume.

This is a potential limitation of the inverse problem posed by nebular spectroscopy, though one that can in principle be overcome with independent physical constraints on the density PDF --- from turbulence models, pressure-balance arguments or resolved observations that restrict the range of physically plausible distributions. We note that the density range of nearly five orders of magnitude required to reproduce the observed hierarchy is much broader than what turbulence alone can sustain in typical H\,\textsc{ii} regions, where ionized density fluctuations already exceed turbulent predictions except in the highest-luminosity sources~\cite{GarciaVazquez:23}. The PDF should therefore be interpreted as an effective distribution encoding the full range of density structures coexisting within the emitting volume --- diffuse gas, dense filaments, ionization fronts and compact condensations --- rather than a single-phase turbulent medium. It does not, however, weaken the central conclusion. Every model consistent with the observations requires a dense component with $\log_{10} n_e \gtrsim 4$~cm$^{-3}$ contributing substantially to the emergent luminosity: a Bayesian fit of the power-law model to the Orion sequence (see Methods) implies that $(10^{+5}_{-4})\%$ of the emission measure --- and hence of the recombination-line emission --- arises from gas above this density, and constrains the distribution to extend to at least $\log_{10} n_e \gtrsim 5.2$ (68\% credibility; $\gtrsim 4.2$ at 95\%) while leaving its dense tail unbounded from above. This is a component that standard low-$n_\mathcal{M}$ diagnostics are largely insensitive to by construction.

Broad, power-law density distributions are not unique to ionized gas. The cold, molecular interstellar medium has long been known to develop extended power-law tails in its density distribution, seen both in the column-density maps of star-forming clouds~\cite{Schneider:15,Chen:18} and in simulations of self-gravitating supersonic turbulence~\cite{FederrathKlessen:13,KlessenGlover:16}, where the tail traces the dense, collapsing gas that forms stars. Our results seem to be consistent with an analogous broad structure in the ionized phase. The slopes are not directly comparable --- cold-ISM studies characterize the volume-weighted distribution of the column density, whereas we constrain the emission-measure-weighted distribution of the volume density --- but recast as an equivalent spherical density profile $\rho \propto r^{-\alpha}$, our slope $\beta \approx -1.3$ corresponds to $\alpha \approx 1.3$ (Methods), somewhat shallower than the $\alpha \approx 1.5$--$2$ typically inferred for molecular clouds~\cite{Schneider:15}. Such a difference may be expected, since photoionization does not map the parent neutral density field one-to-one onto the ionized gas. The breadth of the ionized distribution may therefore be, at least in part, a relic of the gravoturbulent fragmentation of the parent cloud, reprocessed but not erased by the transition to the ionized phase.

Regardless of its origin, this dense component --- substantial in luminosity yet small in volume --- has consequences that extend well beyond the density measurements themselves. Pioneering works~\cite{Panagia:75,Aller:76,Rubin:89} already warned that density inhomogeneities introduce systematic biases into nebular diagnostics, but the universality and predictability of those biases, their dependence on the atomic response functions of the diagnostics, and the observational evidence now establishing them across such a broad range of environments, had not previously been demonstrated. The present results show that every physical quantity derived from forbidden-line spectroscopy under the single-density approximation inherits a bias whose magnitude and direction depend on which diagnostic is used.

An immediate consequence concerns the thermal structure of the nebula. In regions where the electron density approaches or exceeds the critical density of the dominant coolants (e.g. [O\,\textsc{iii}]~$\lambda5007$, [O\,\textsc{ii}]~$\lambda\lambda3727,3729$), collisional de-excitation suppresses forbidden-line cooling, and the gas reaches a higher thermal equilibrium temperature than the surrounding diffuse medium. A broad density PDF therefore may not merely bias density measurements --- it can introduce genuine temperature inhomogeneities as a direct physical consequence of the density structure itself. Density inhomogeneities are already known to distort auroral-to-nebular temperature and abundance diagnostics~\cite{Viegas:94}, and, when combined with grain heating, to modify the thermal balance of the gas~\cite{Stasinska:01}. The temperature inhomogeneities quantified by $t^2$~\cite{Peimbert:67,MendezDelgado:2023a} may therefore be, at least in part, a physical imprint of unresolved density distributions rather than an independent phenomenon.

The implications propagate further. Ionized gas masses require the volume-weighted mean density, but forbidden-line diagnostics return emissivity-weighted densities biased toward the densest, most luminous structures; gas masses derived from [S\,\textsc{ii}] are therefore likely biased. ISM pressures inferred from high-$n_\mathcal{M}$ diagnostics may overestimate the pressure of the bulk gas. Filling factors, which require both a volume-weighted density and an emission measure, are doubly compromised. Chemical abundances derived from collisionally excited lines depend on both temperature and density at every step, and both are now known to be biased in ways that depend on the diagnostic used. Galaxy feedback energetics, outflow masses and momentum fluxes --- quantities central to our understanding of galaxy evolution --- all rest on density estimates that are subject to the bias quantified here.

The standard framework of nebular spectroscopy --- constant-density photoionization models, single representative densities, temperature diagnostics applied without regard to the density regime they sample --- is therefore not merely incomplete. It is systematically biased in a predictable way. The slope of equation~(\ref{eq:orion_fit}) is the fingerprint of that bias. Correcting for it requires abandoning the single-density approximation and replacing it with models that incorporate realistic density distributions --- and exploiting the diagnostic hierarchy itself as the primary observational route to recovering the unresolved density structure of ionized gas across the Universe.

%% ============================================================
%% METHODS
%% ============================================================

\section*{Methods}

\subsection*{LVM observations of the Orion Nebula}

The Orion Nebula data were obtained as part of the SDSS-V Local Volume Mapper (LVM) survey~\cite{Herbst:2024,Drory:2024,Sanchez:2025,Kollmeier:2026}, which provides integral-field spectroscopy with wavelength coverage $\sim$3600--9800~\AA, spectral resolution $R \sim 4000$ and a fibre diameter of $35.3''$, corresponding to $\sim$0.07~pc at the adopted distance of 410~pc~\cite{MendezDelgado:2022}. The science field of view comprises 1801 fibres arranged in a hexagonal pattern. Eight long-exposure frames of 900~s each were combined to cover the full M42--M43 region (exposure IDs: 8487, 9478, 8909, 9578, 9545, 8911, 8823, 8822). Short-exposure frames of 10~s each (exposure IDs: 8254, 10151, 10153, 10154) were obtained at the same pointings to enable saturation correction of the brightest lines. The raw data were reduced with the standard LVM data-reduction pipeline (version 1.2.1; A.~Mej\'{i}a-Narv\'{a}ez et al., in preparation), which performs detector detrending, wavelength calibration from arc-lamp exposures, flux calibration from simultaneous standard-star observations, and sky subtraction using dedicated sky fibres. Three-dimensional data cubes were reconstructed from the row-stacked spectra using the \textsc{3DCubeGen} package (H.~Ibarra-Medel et al., in preparation) following the spatial interpolation scheme described in \cite{Orozco-Duarte:26}.

In the Huygens region of M42, the following lines are saturated in the 900~s long-exposure data: H\,\textsc{i}~$\lambda4861$ (H$\beta$), H\,\textsc{i}~$\lambda6563$ (H$\alpha$), [N\,\textsc{ii}]~$\lambda6548$, [N\,\textsc{ii}]~$\lambda6583$, [O\,\textsc{iii}]~$\lambda4959$, [O\,\textsc{iii}]~$\lambda5007$, [S\,\textsc{iii}]~$\lambda9069$ and [S\,\textsc{iii}]~$\lambda9531$. For each saturated spaxel, the line profile was replaced by a renormalized version of the corresponding 10~s short-exposure spectrum. The two exposures were aligned to a common wavelength solution through a correction derived from the median offset of unsaturated lines detected at sufficient signal-to-noise in both, and the short-exposure spectrum was renormalized to the long-exposure continuum level via a linear fit to adjacent spectral windows. 

\subsection*{Emission-line measurements}

Emission-line fluxes were measured fibre-by-fibre from the reconstructed data cubes using a custom line-fitting described in ~\cite{Sattler:2026}. For each emission line, the local continuum was modeled by a linear function fitted simultaneously to two spectral windows on either side of the line, and a single Gaussian profile was fitted. In spectral regions affected by broad stellar absorption features underlying the nebular emission --- most notably near the Balmer lines --- the continuum model was augmented by an additional Gaussian absorption component, whose parameters were constrained by the continuum windows and validated against internal stability criteria; if the absorption solution was found to be unphysical or numerically unstable, the code reverted to a purely linear continuum. Multiple initial guesses for the Gaussian width were attempted at each spaxel to ensure convergence, and fits were accepted only when the resulting integrated flux, full-width at half maximum and signal-to-noise ratio satisfied internal quality criteria. Flux uncertainties were propagated from the rms noise of the continuum residuals as $\sigma_F = \sigma_\mathrm{cont}\,\delta\lambda\,\sqrt{N_\mathrm{pix}}$, where $\delta\lambda$ is the pixel width and $N_\mathrm{pix}$ is the number of pixels in the line window. The partially blended [O\,\textsc{ii}]~$\lambda\lambda3726,3729$, [N\,\textsc{i}]~$\lambda\lambda5198,5200$ doublets were deblended by simultaneously fitting two Gaussian components with wavelength centroids constrained by atomic physics. All spaxels with $S/N < 3$ in any diagnostic line used in subsequent analysis were masked.

\subsection*{Reddening correction}

Interstellar extinction was corrected using the \textsc{PyNeb} \texttt{RedCorr} module~\cite{Luridiana:15} with the \cite{Cardelli:89} reddening law optimized for the Orion Nebula by \cite{Blagrave:07} and $R_V = 5.5$. The extinction coefficient $c(\mathrm{H}\beta)$ was computed at each spaxel as the median of eight independent estimates from pairs of H\,\textsc{i} recombination lines: H$\alpha$/H$\beta$, H$\gamma$/H$\beta$, H$\delta$/H$\beta$, H$\gamma$/P9, H$\beta$/P9, H$\delta$/P9, H$\alpha$/P9 and H$\delta$/H$\alpha$, where P9 refers to Paschen-9 ($\lambda9229$). Theoretical ratios were computed from the effective recombination coefficients of \cite{Storey:95} at $T_e = 9000 \pm 2000$~K using the pre-computed [S\,\textsc{ii}] density map as an initial estimate. The use of multiple line pairs spanning the full LVM wavelength range substantially reduces the sensitivity of the final $c(\mathrm{H}\beta)$ map to systematic effects in any individual ratio.

\subsection*{Electron density maps}

Electron densities were derived from the following 16 forbidden-line diagnostics: [Fe\,\textsc{ii}]~$\lambda8617/\lambda7155$; [S\,\textsc{ii}]~$\lambda6717/\lambda6731$, $\lambda4070/\lambda6717$, $\lambda4070/\lambda6731$, $\lambda4070/(\lambda6717+\lambda6731)$; [O\,\textsc{ii}]~$\lambda3727/\lambda3729$, $\lambda(\lambda7319+7320)/\lambda3727$, $\lambda(\lambda7319+7320)/\lambda3729$, $(\lambda\lambda7319+7320)/(\lambda3727+\lambda3729)$; [Fe\,\textsc{iii}]~$\lambda4658/\lambda4986$, $\lambda4658/\lambda4702$, $\lambda4658/\lambda4881$, $\lambda4986/\lambda4702$, $\lambda4986/\lambda4881$, $\lambda4702/\lambda4881$; and [Cl\,\textsc{iii}]~$\lambda5538/\lambda5518$. The [O\,\textsc{ii}]~$\lambda\lambda7330,7331$ lines were not used owing to contamination by sky emission, and the [S\,\textsc{ii}]~$\lambda4076$ line was excluded because its smaller emitting area relative to $\lambda4070$ results in substantially reduced spatial coverage at $S/N > 3$. Special care was taken to identify and flag potential sky-line contamination throughout the nebular field. Although available in principle, the [N\,\textsc{i}]~$\lambda5198/\lambda5200$ diagnostic was excluded from the linear fit of equation~(\ref{eq:orion_fit}) for two reasons: residual sky contamination in this spectral region cannot be fully ruled out, and the [N\,\textsc{i}] doublet is known to receive a non-negligible contribution from continuum fluorescence in addition to collisional excitation~\cite{Ferland:2012}, violating the purely collisional assumption underlying the density inversion. The [Fe\,\textsc{ii}] transitions at $\lambda8617$ and $\lambda7155$ are not affected by the fluorescent pumping that operates through the metastable $a\,^4F_{9/2}$ level~\cite{Rodriguez:99,Verner:2000}, as confirmed by several studies~\cite{Baldwin:1996,MendezDelgado:21,Mendoza:23}.

For each diagnostic, densities were computed spaxel-by-spaxel using the \textsc{PyNeb} \texttt{getTemDen} function~\cite{Luridiana:15} via its machine-learning implementation in \textsc{ai4neb}. Uncertainties were propagated through 1000 Monte Carlo realizations per spaxel, perturbing the line fluxes according to their Gaussian error distributions and the electron temperature according to $T_e = 9000 \pm 2000$~K.

\begin{figure*}
\centering
\begin{tabular}{ccc}
% Row 1: log nM = 2.91, 3.06, 3.16
\includegraphics[width=0.31\textwidth]{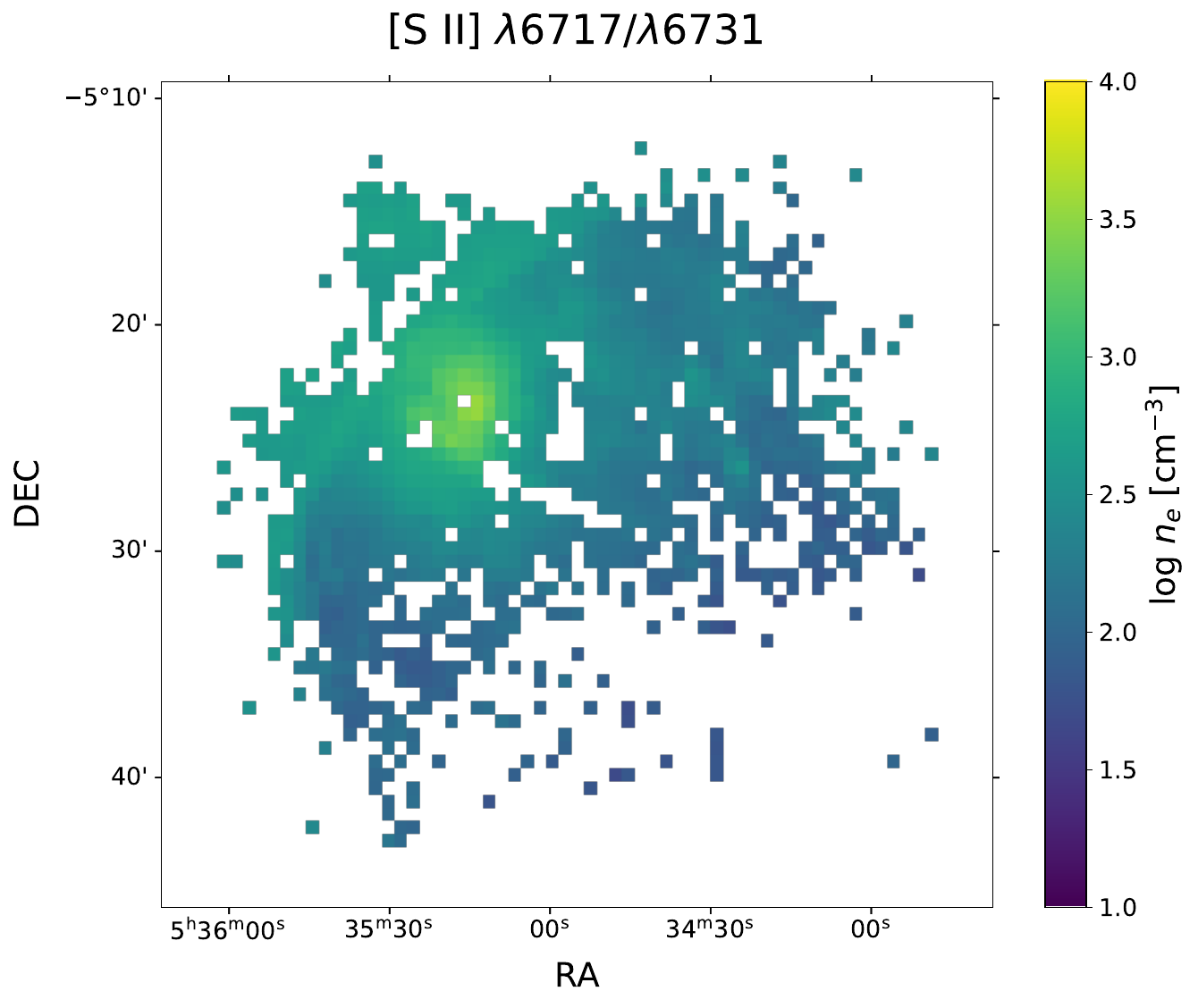} &
\includegraphics[width=0.31\textwidth]{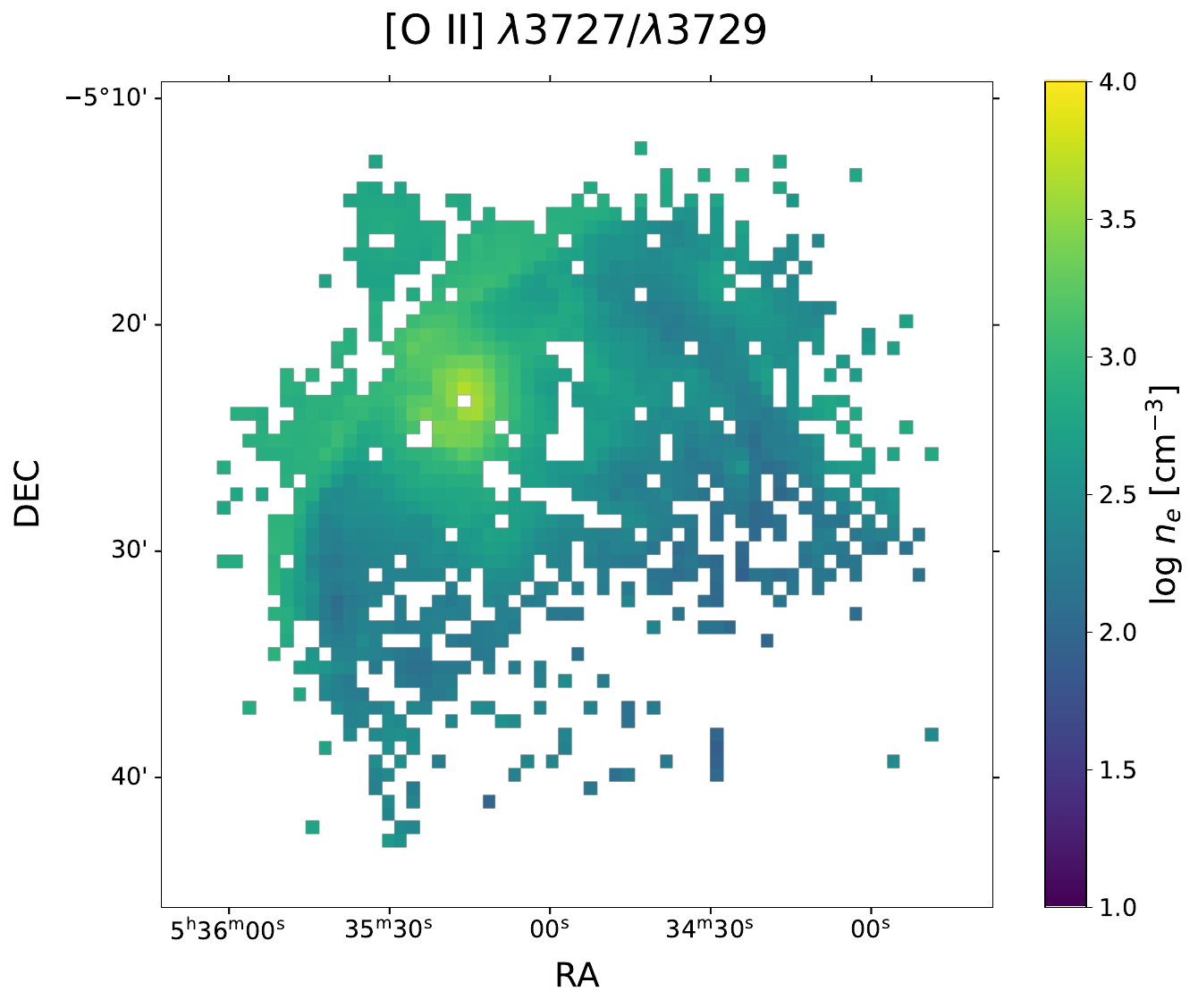} &
\includegraphics[width=0.31\textwidth]{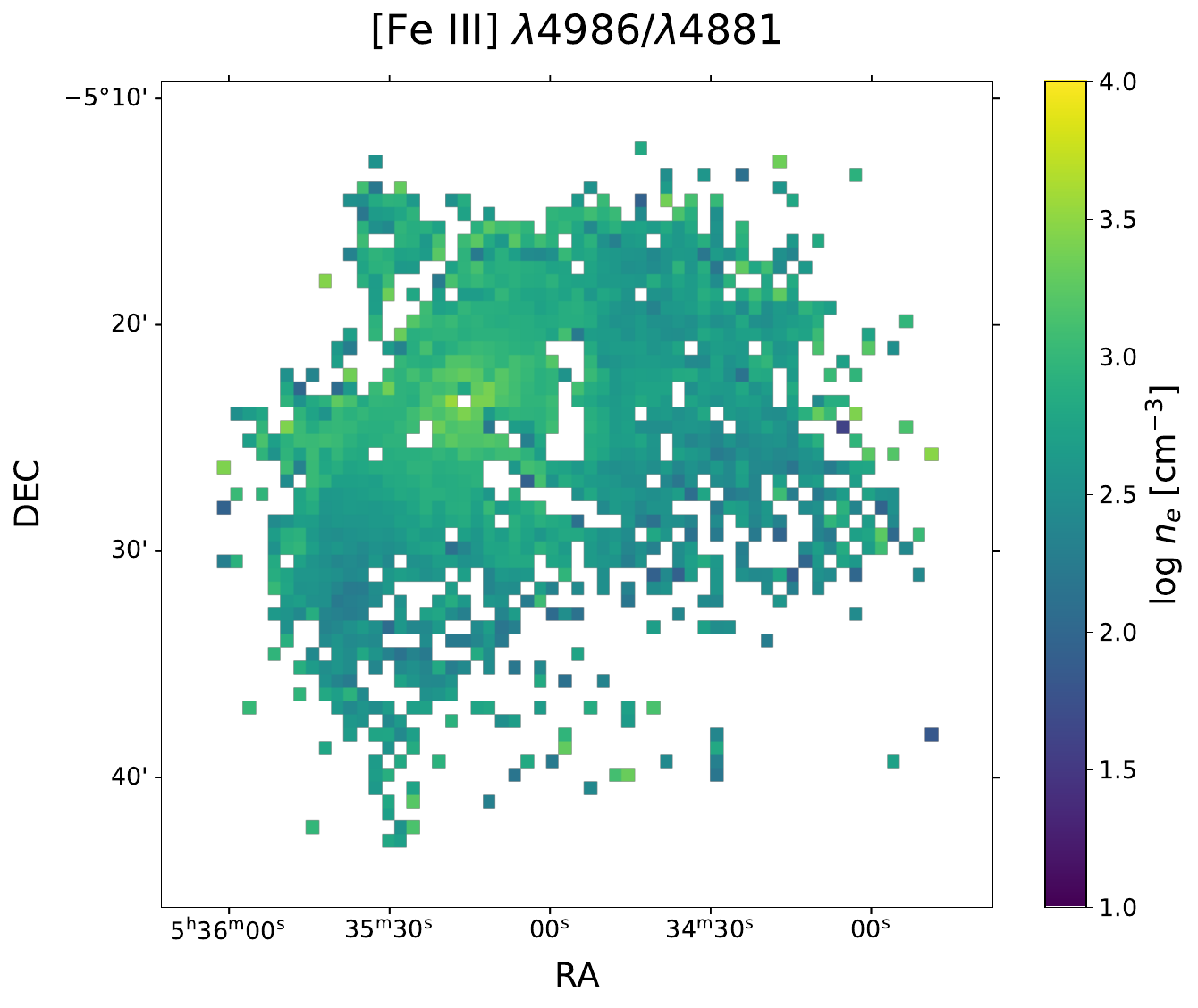} \\
% Row 2: log nM = 3.74, 3.84, 3.99
\includegraphics[width=0.31\textwidth]{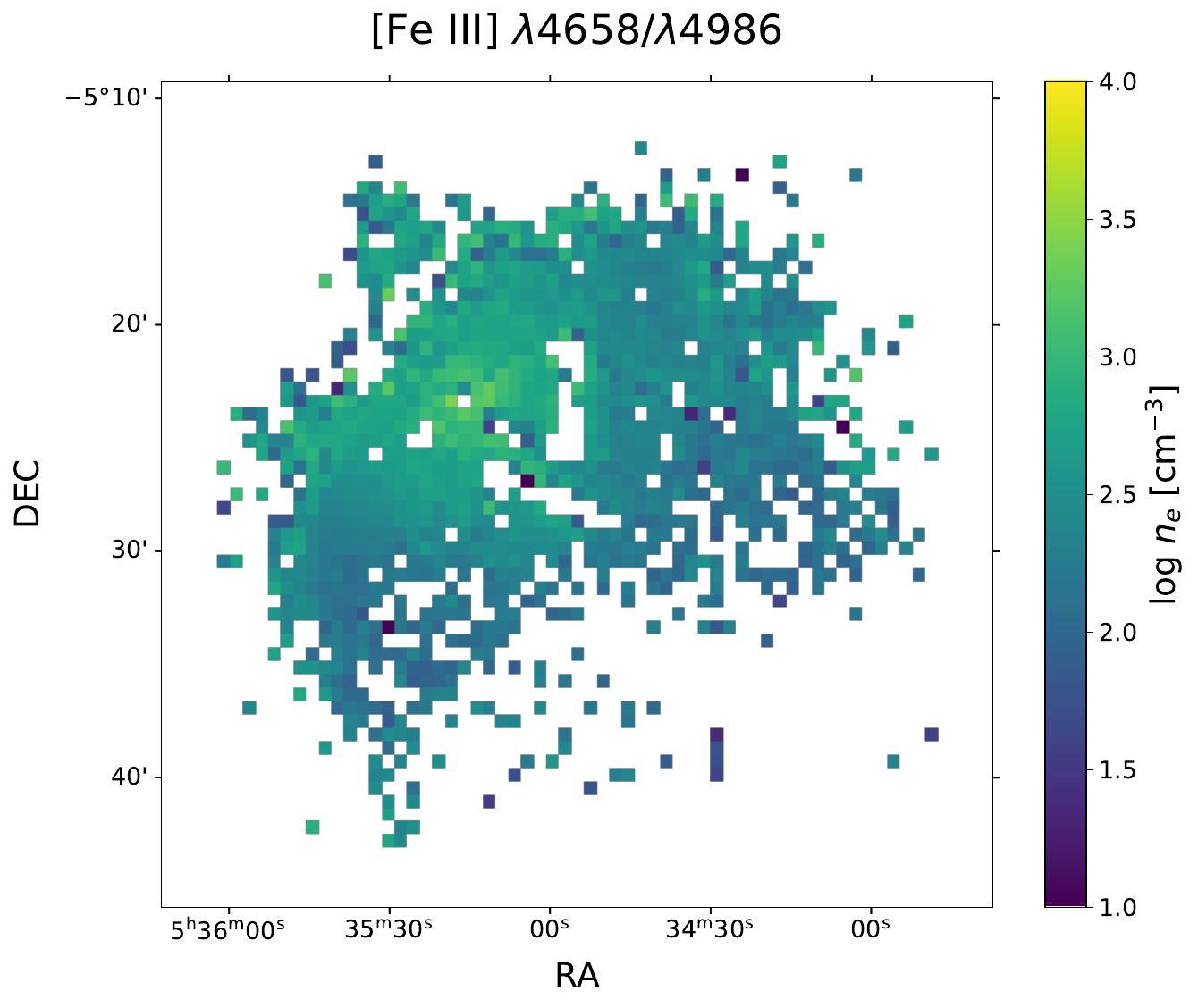} &
\includegraphics[width=0.31\textwidth]{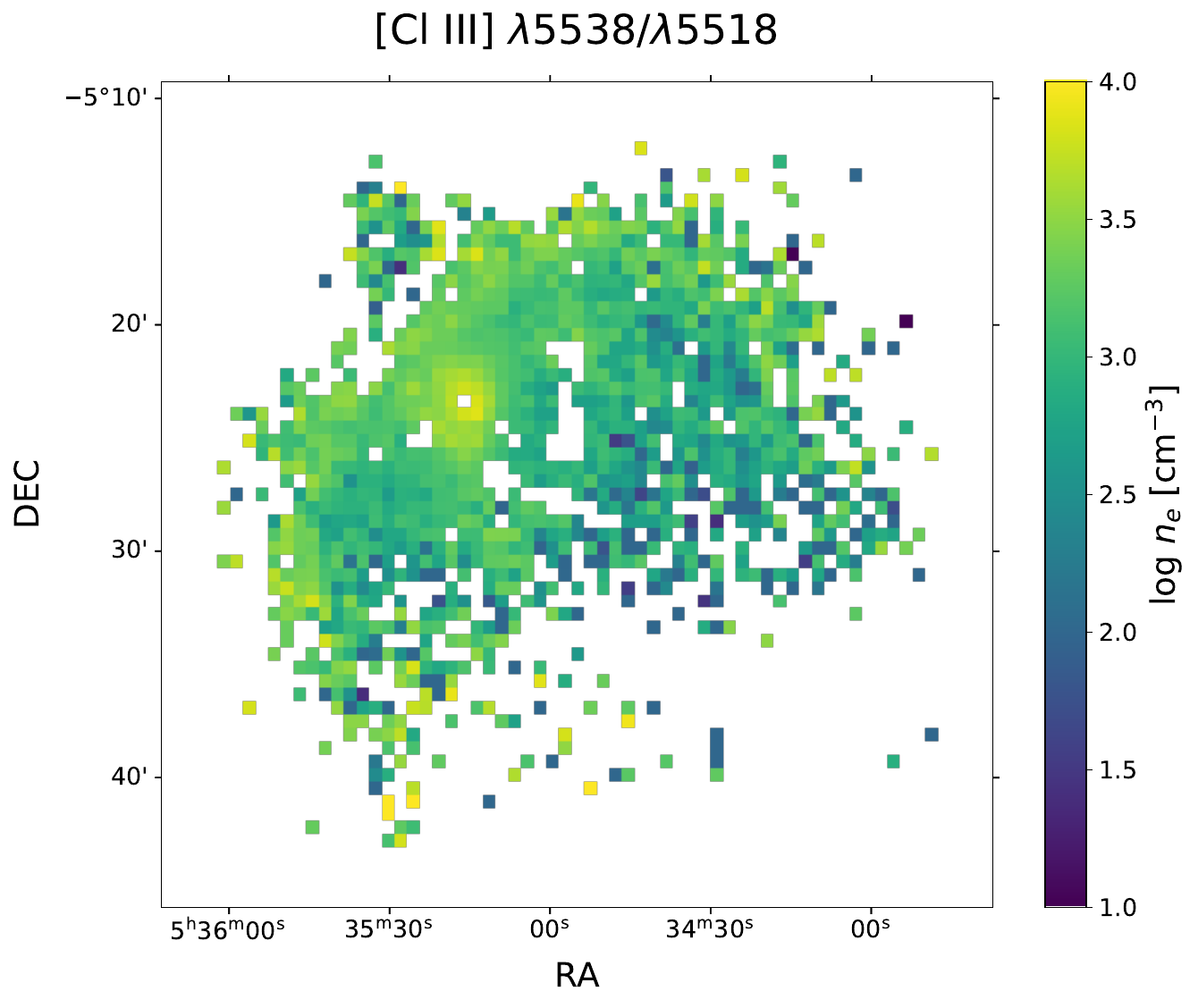} &
\includegraphics[width=0.31\textwidth]{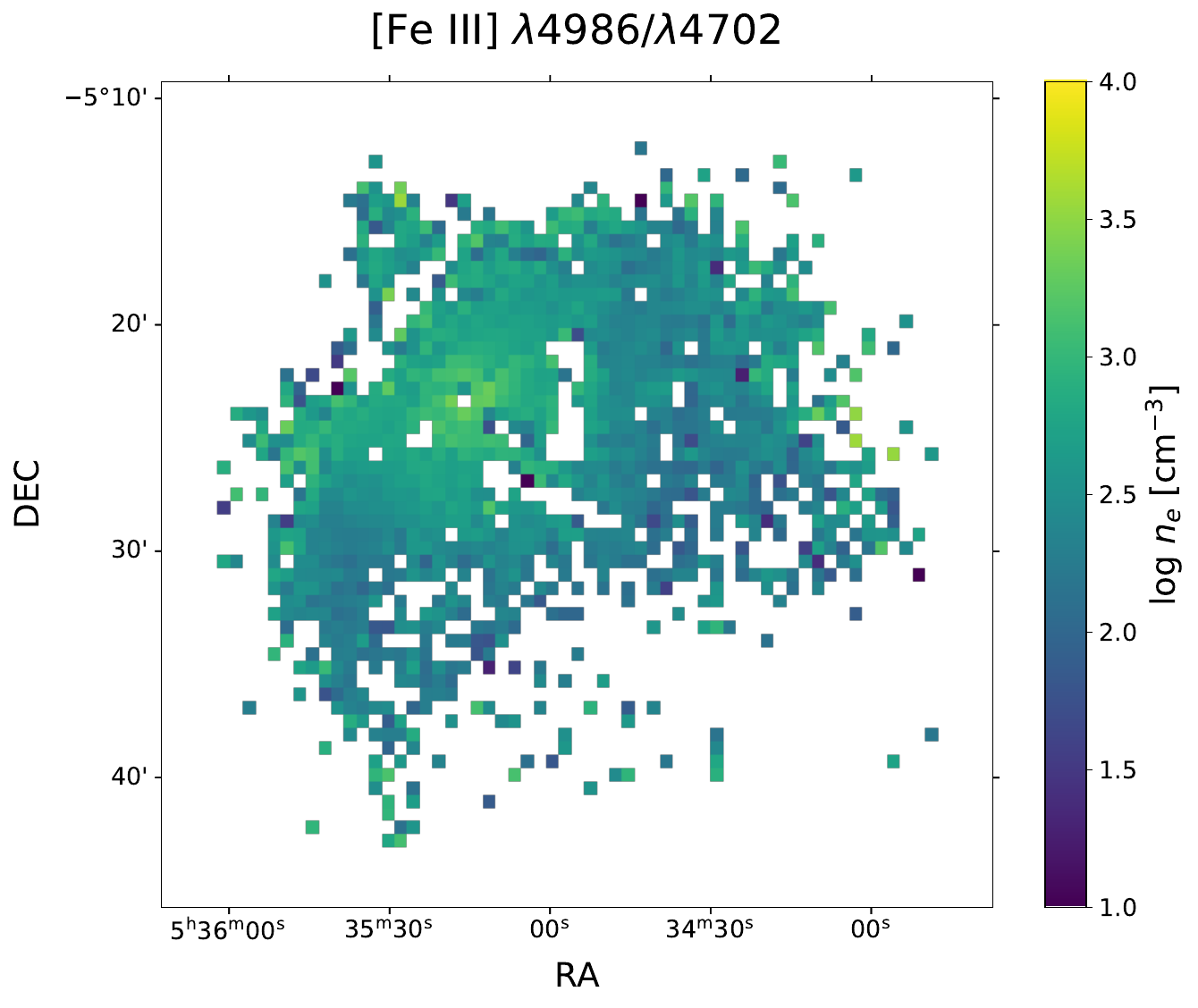} \\
% Row 3: log nM = 4.04, 4.64, 4.69
\includegraphics[width=0.31\textwidth]{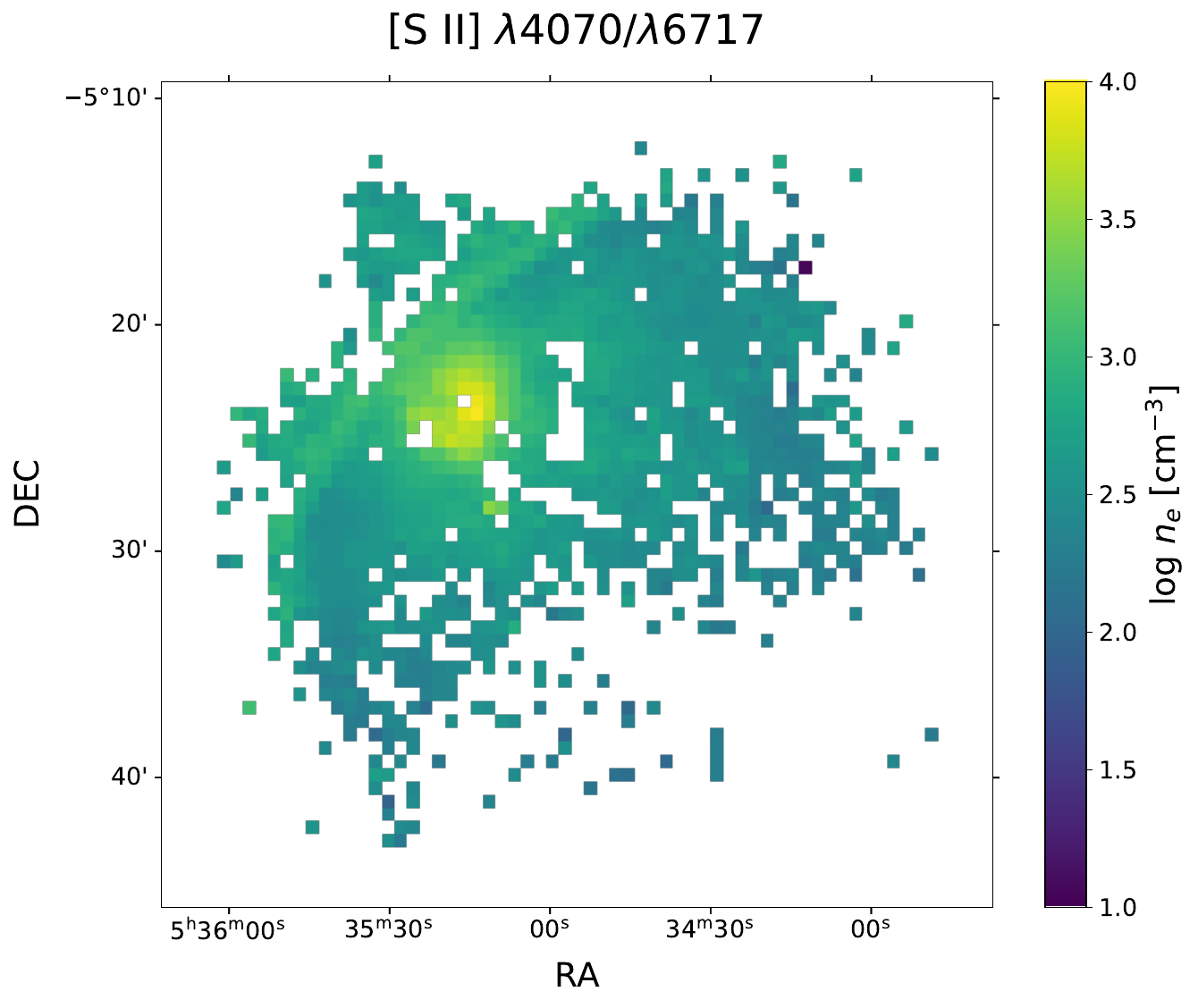} &
\includegraphics[width=0.31\textwidth]{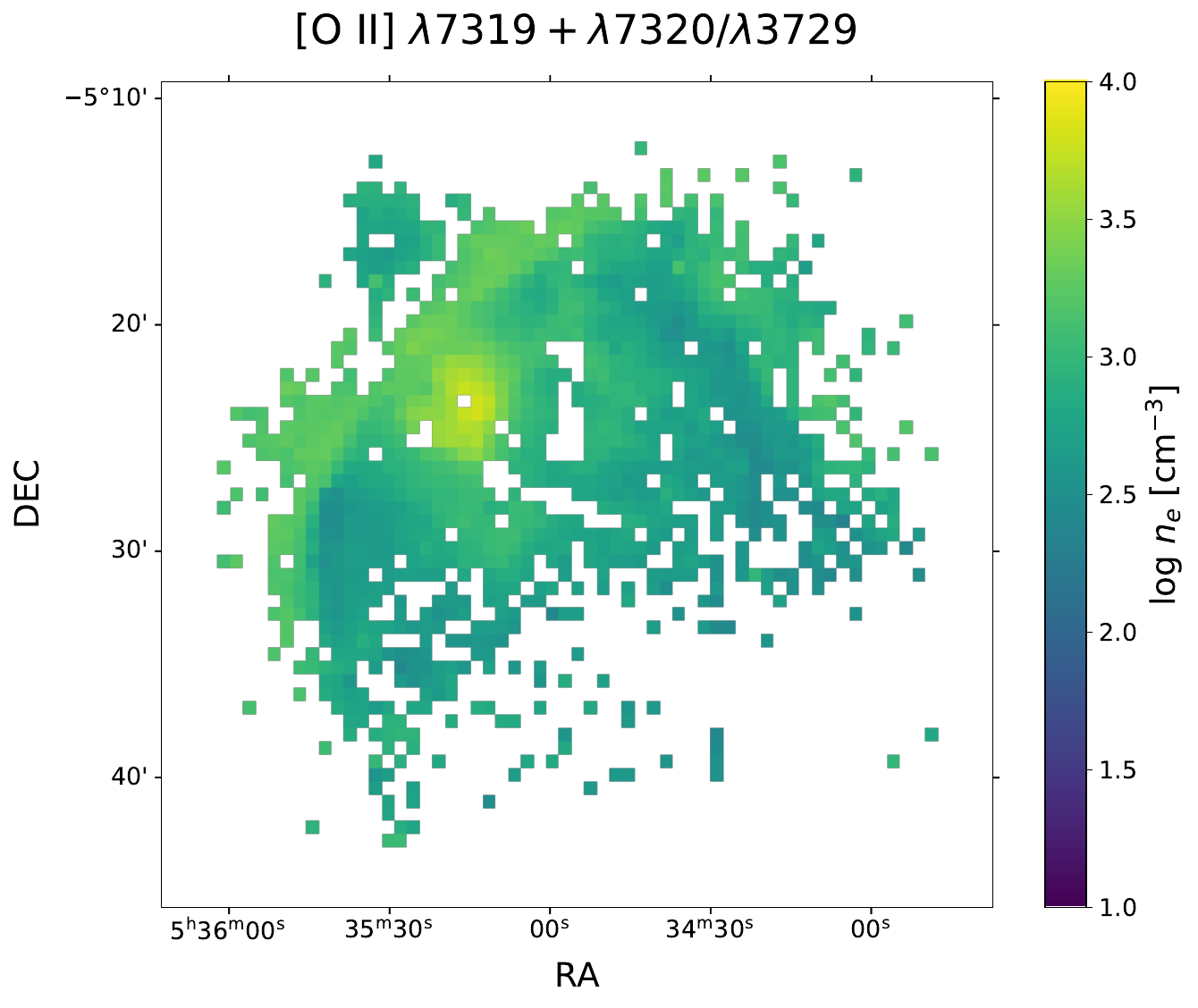} &
\includegraphics[width=0.31\textwidth]{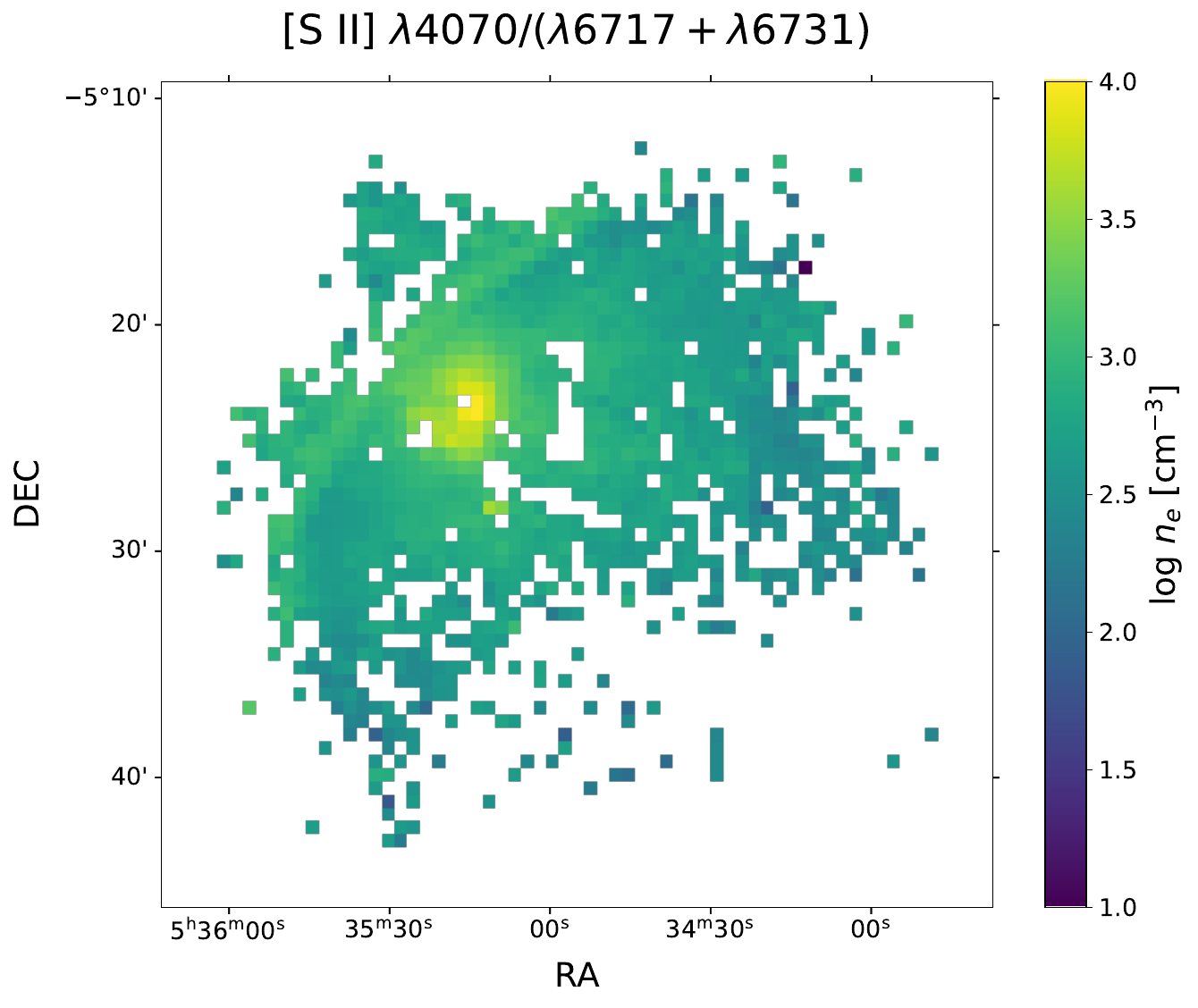} \\
% Row 4: log nM = 4.85, 4.91, 5.03
\includegraphics[width=0.31\textwidth]{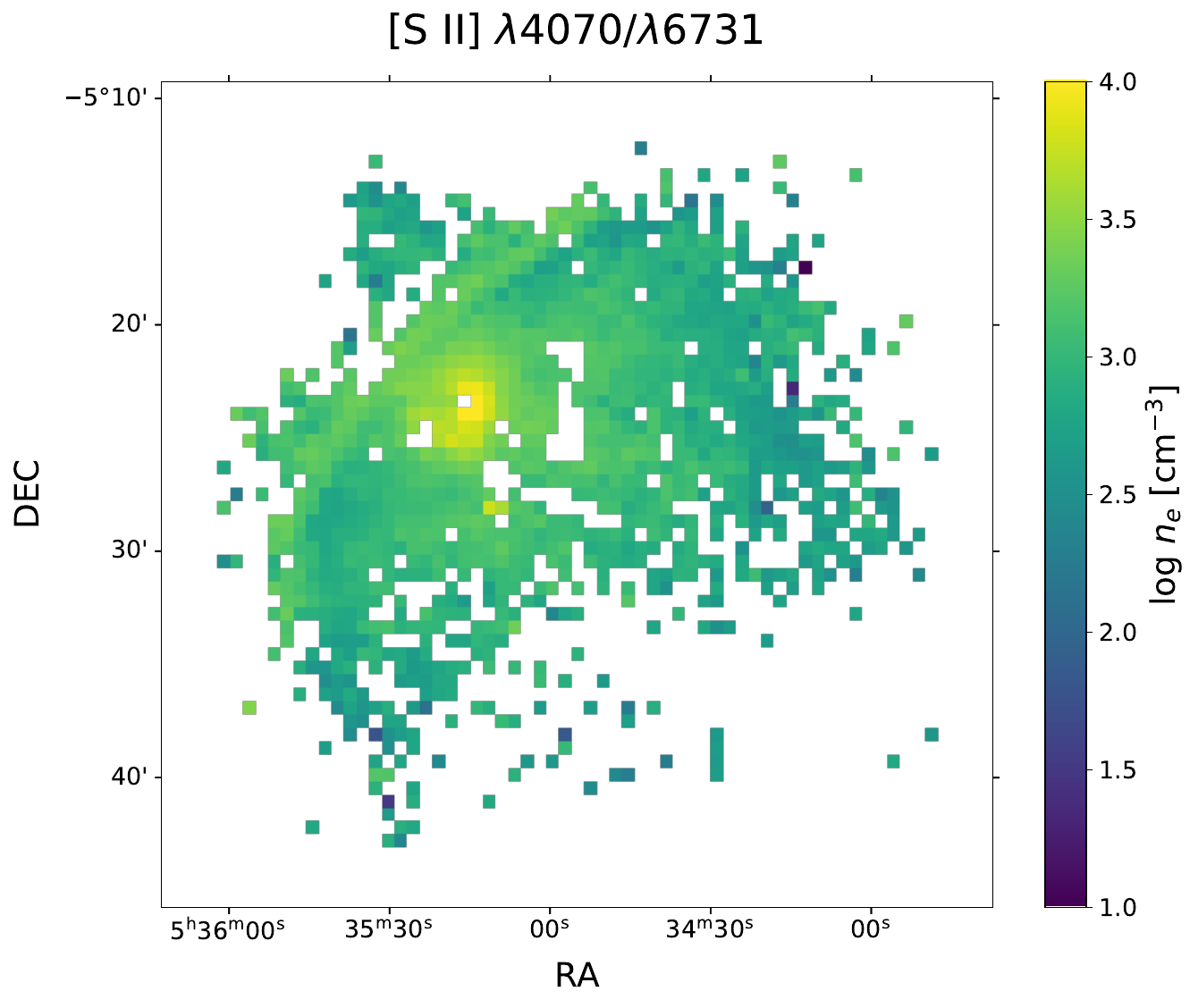} &
\includegraphics[width=0.31\textwidth]{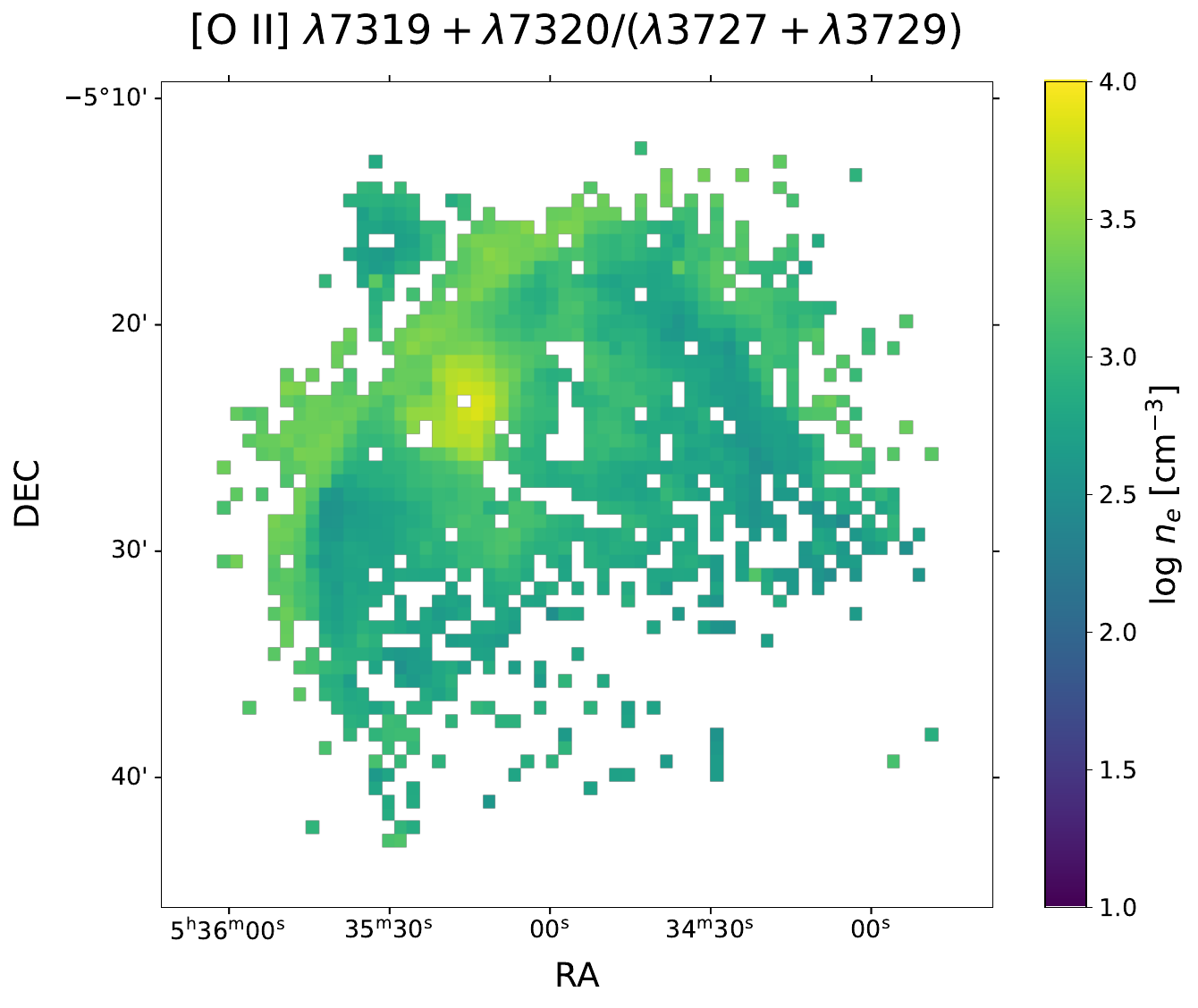} &
\includegraphics[width=0.31\textwidth]{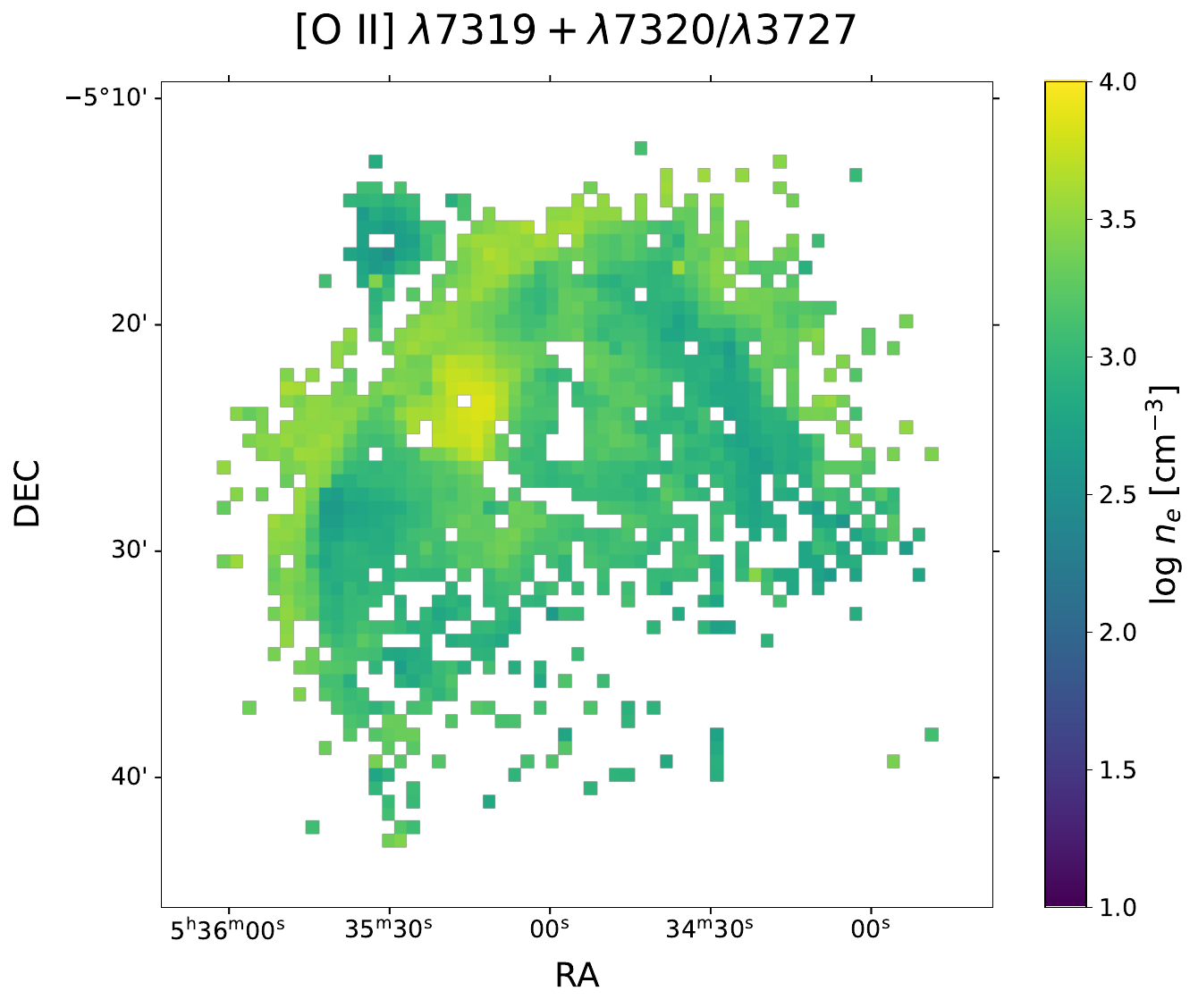} \\
% Row 5: log nM = 5.10, 6.19, 6.31
\includegraphics[width=0.31\textwidth]{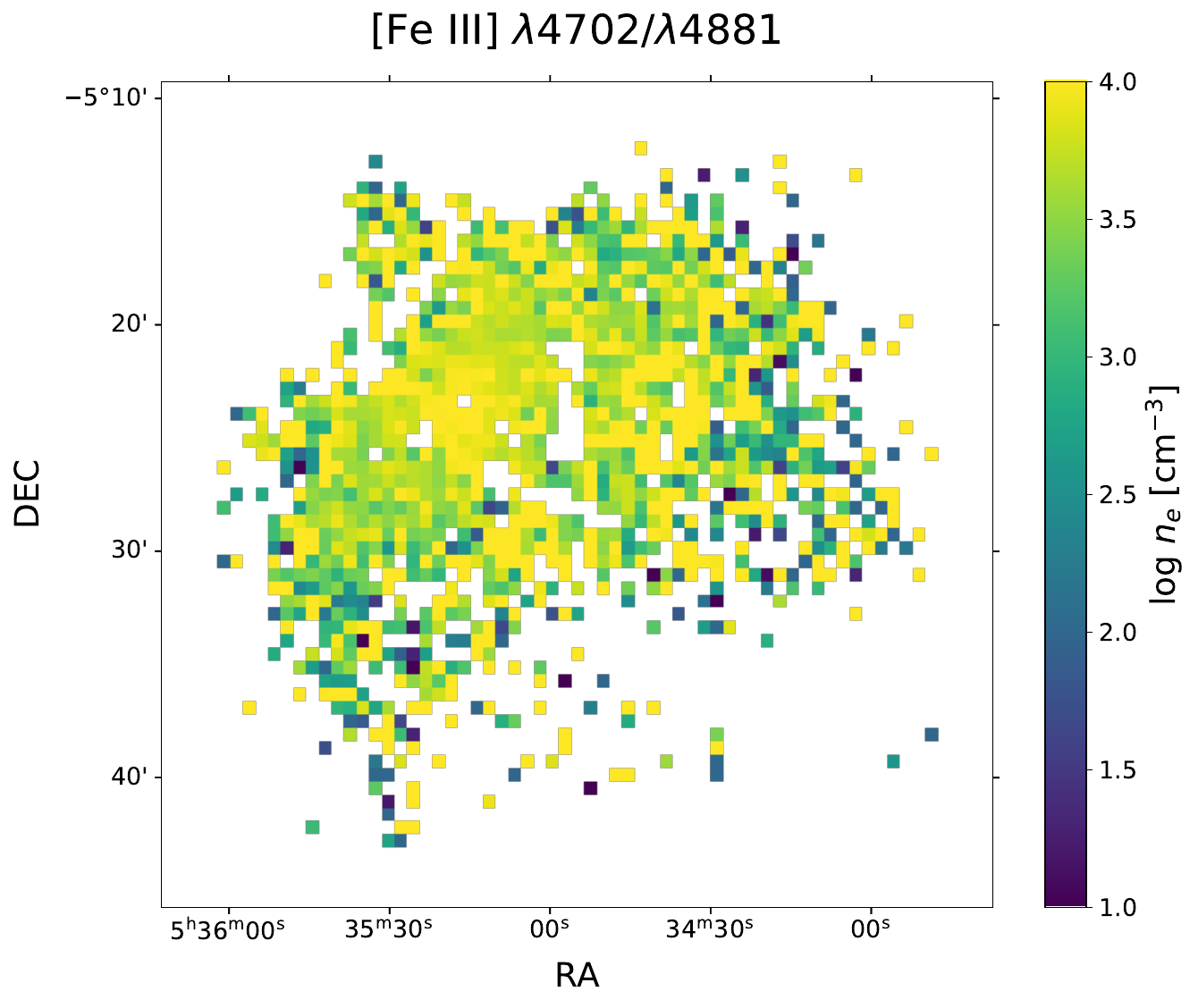} &
\includegraphics[width=0.31\textwidth]{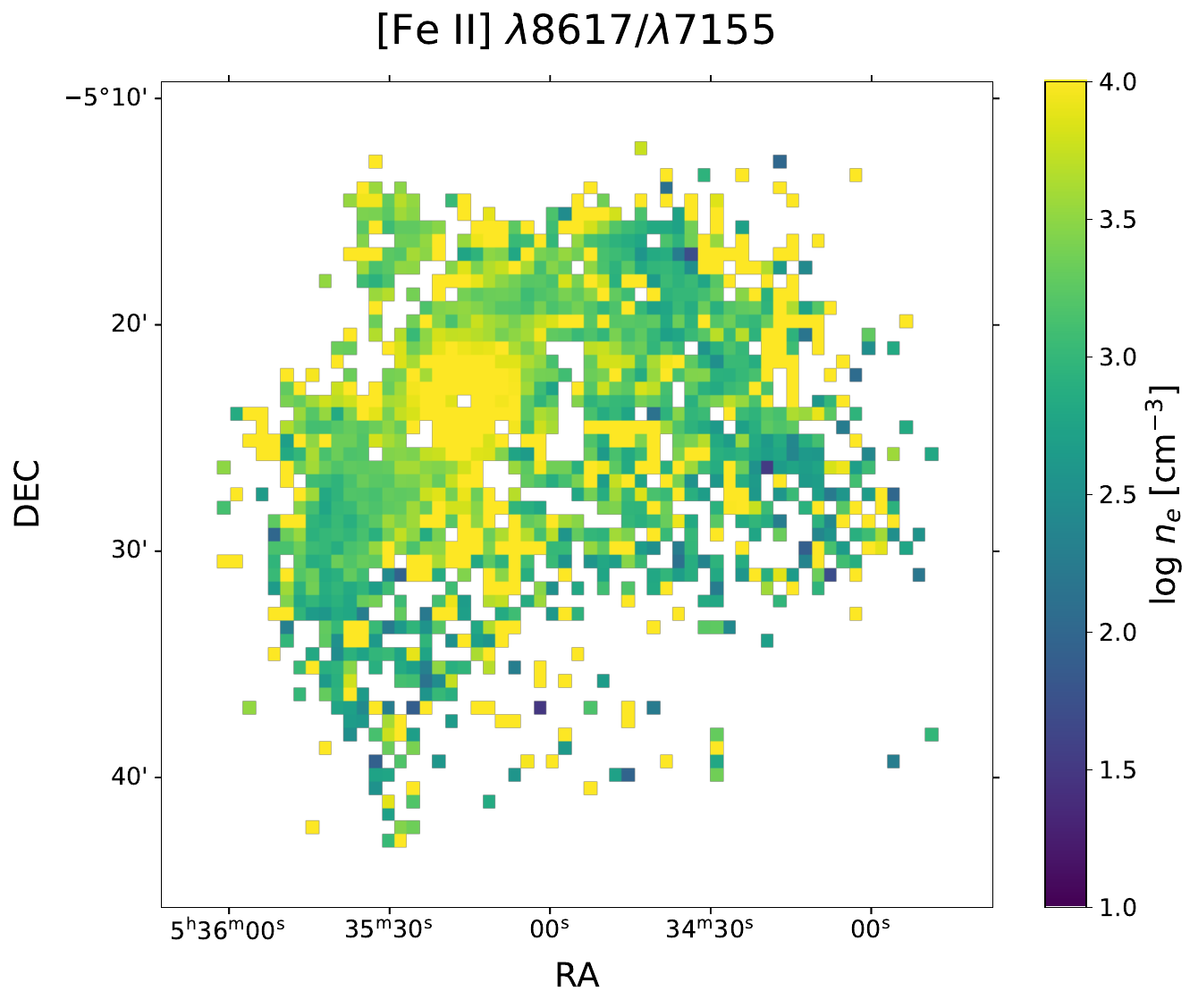} &
\includegraphics[width=0.31\textwidth]{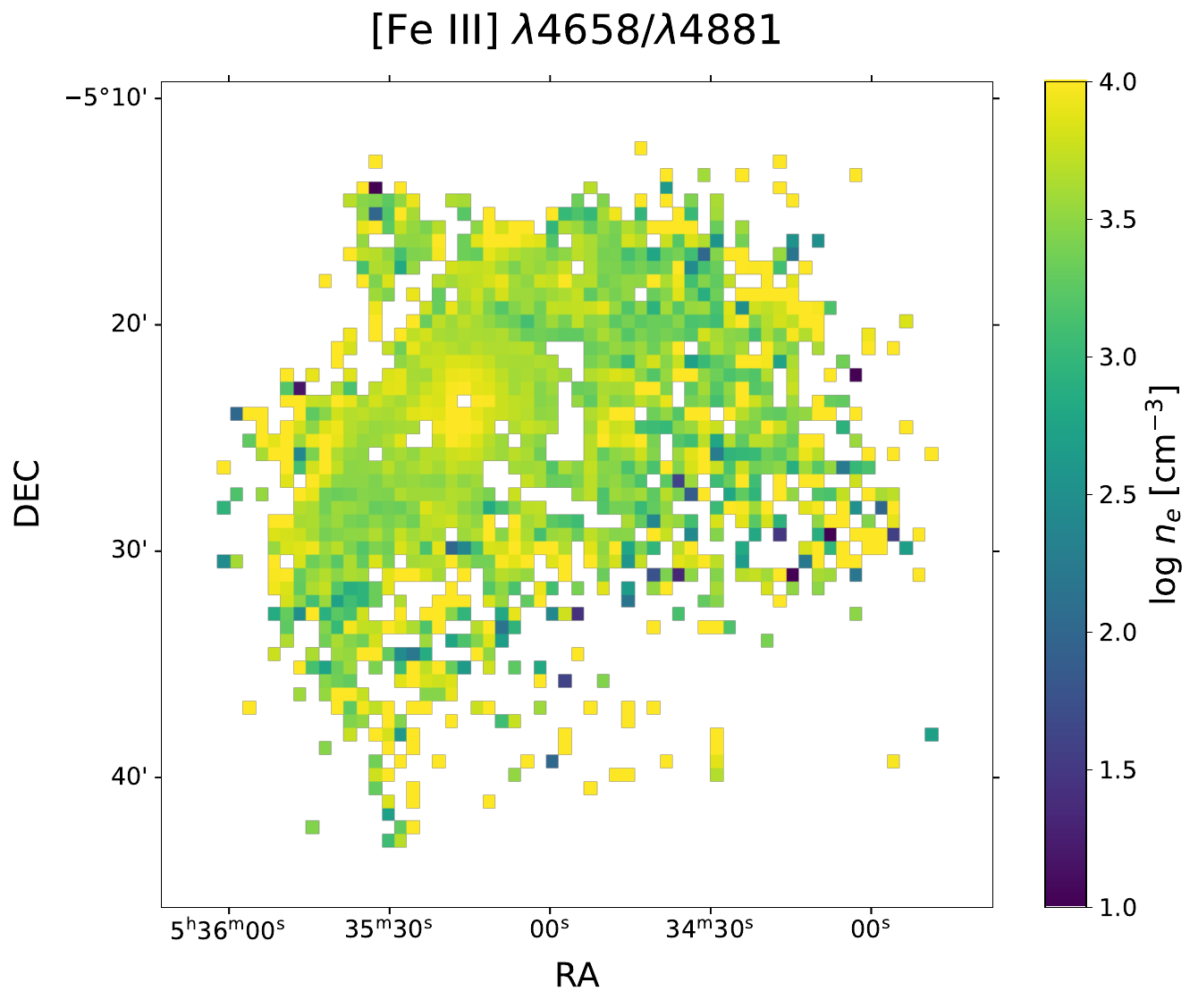} \\
% Row 6: log nM = 6.36, 6.90
\includegraphics[width=0.31\textwidth]{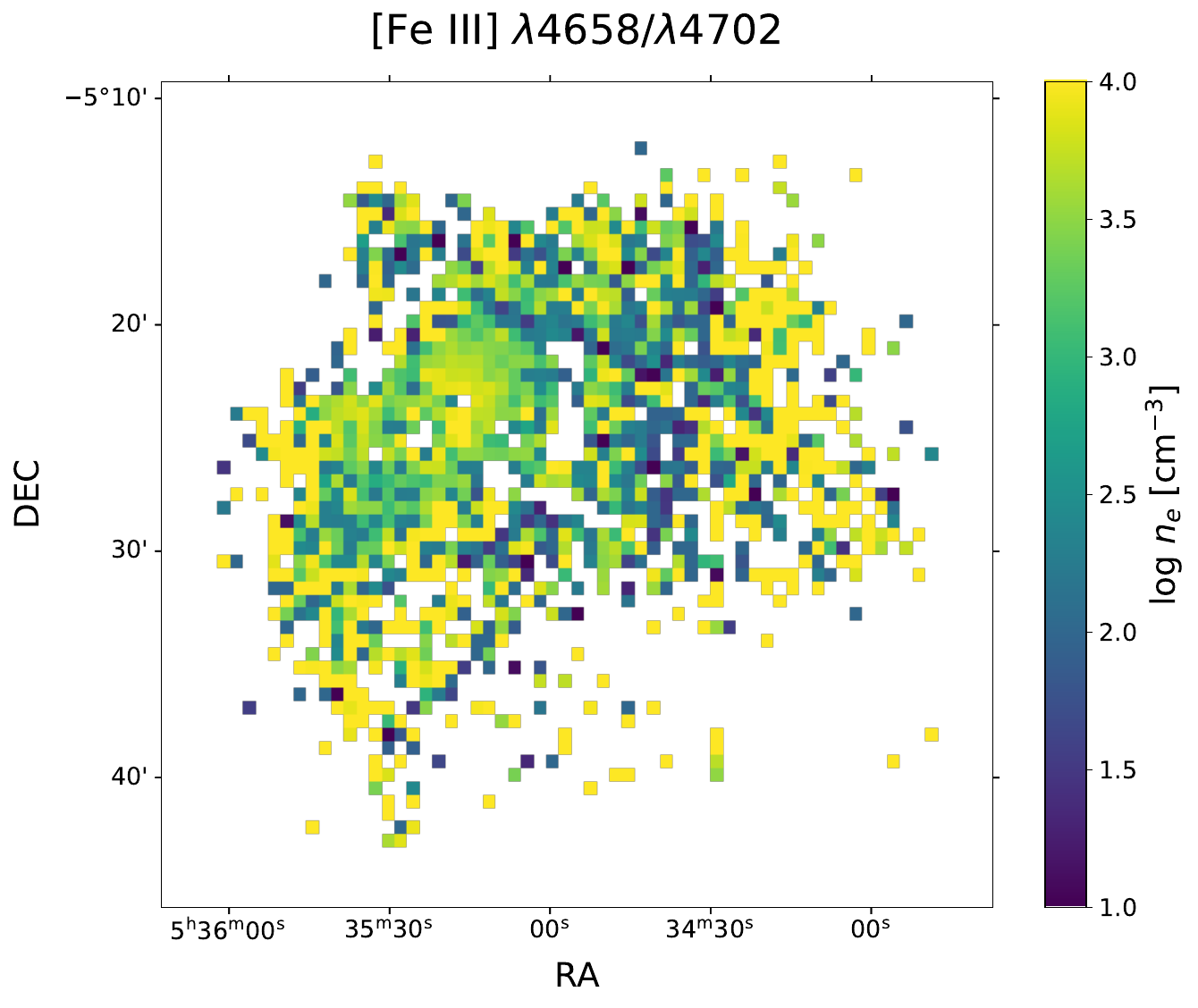} &
\phantom{\includegraphics[width=0.31\textwidth]{map_neFe3_4658_4702.pdf}} \\
\end{tabular}
\caption{\textbf{Electron density maps from all 16 diagnostics in the common region of the Orion Nebula.}
Spaxel-by-spaxel electron density maps derived from each of the 16 forbidden-line diagnostics used in this study, restricted to the common M42--M43 region of 1226 spaxels in which all diagnostics are simultaneously detected at $S/N > 3$, and ordered by increasing maximum-sensitivity density $n_\mathcal{M}$ (values given in Extended Data Table~1).  All panels share the same colour scale in units of $\log\,n_e$~[cm$^{-3}$].}
\label{fig:extdata_maps}
\end{figure*}

For diagnostics involving auroral-to-nebular line ratios of [S\,\textsc{ii}] ($\lambda4070/\lambda6717$, $\lambda4070/\lambda6731$, $\lambda4070/(\lambda6717+\lambda6731)$) and [O\,\textsc{ii}] (($\lambda\lambda7319+7320)/\lambda3727$, $(\lambda\lambda7319+7320)/\lambda3729$, $(\lambda\lambda7319+7320)/(\lambda3727+\lambda3729)$), the electron temperature adopted in the density inversion was that measured from [N\,\textsc{ii}]~$\lambda5755/\lambda6584$ rather than the nominal 9000~K. The ions involved --- S$^+$, O$^+$ and N$^+$ --- predominantly coexist in the same low-ionization zone of the nebula, making $T_e(\text{[N\,\textsc{ii}]})$ the physically appropriate temperature for these diagnostics~\cite{Rickards:2024,MendezDelgado:23b}. This ensures that within-ion comparisons between auroral-to-nebular and classical doublet ratios of the same species are performed on equal thermodynamic footing. The common region where all 16 diagnostics are simultaneously detected at $S/N > 3$ comprises 1226 spaxels. Of these 16 diagnostics only nine are algebraically independent: for an ion with $k$ density-sensitive lines, only $k-1$ line ratios are mutually independent, the remainder being exact combinations of these. The independent set comprises two S$^+$ ratios, two O$^+$, three Fe$^{2+}$, one Cl$^{2+}$ and one Fe$^+$. The linear fit of equation~(\ref{eq:orion_fit}) is performed on these nine ratios only; including all 16 returns a consistent slope and correlation but an over-optimistic significance ($p=3.9\times10^{-6}$ instead of $8.6\times10^{-4}$), because the dependent ratios carry no independent statistical information. The fitted slope remains intermediate between 0 and 1 ($\approx 0.29$--$0.33$) across alternative choices of the independent set, so the inferred hierarchy does not depend on the particular ratios adopted. All 16 ratios are shown in Fig.~\ref{fig:main_sequence}. Electron density maps are shown in Extended Data Fig.~\ref{fig:extdata_maps}.

\subsection*{DESIRED sample}

As part of the DESIRED (\textit{DEep Spectra of Ionised Regions Database}) project~\cite{MendezDelgado:23b}, we compile all reported emission-line intensities from Galactic and extragalactic H\,\textsc{ii} regions, star-forming galaxies, and high-redshift systems available in the literature with direct determinations of electron temperature. The database assembles published deep optical spectra of ionized nebulae with high signal-to-noise detections of faint auroral lines, enabling direct determinations of physical conditions via the standard CEL method. High-redshift star-forming galaxies spanning $z \approx 1.6$--10.2 are treated as a separate subsample throughout this analysis.

Physical conditions for each spectrum were computed homogeneously with \textsc{PyNeb}~\cite{Luridiana:15} following the iterative temperature--density procedure described in ~\cite{MendezDelgado:23b}. Electron densities were derived from all available forbidden-line diagnostics: [S\,\textsc{ii}]~$\lambda6717/\lambda6731$, [O\,\textsc{ii}]~$\lambda3727/\lambda3729$, [Cl\,\textsc{iii}]~$\lambda5538/\lambda5518$, [Ar\,\textsc{iv}]~$\lambda4740/\lambda4711$, [Fe\,\textsc{iii}]~$\lambda4658/\lambda4986$, $\lambda4658/\lambda4702$, $\lambda4658/\lambda4881$, $\lambda4986/\lambda4702$, $\lambda4986/\lambda4881$, $\lambda4702/\lambda4881$, [O\,\textsc{ii}]~$(\lambda\lambda7319+7320+7330+7331)/(\lambda3727+\lambda3729)$, [S\,\textsc{ii}]~$(\lambda4070+\lambda4075)/(\lambda6717+\lambda6731)$, C\,\textsc{iii}]~$\lambda1909/\lambda1907$, Si\,\textsc{iii}]~$\lambda1882/\lambda1892$ and [N\,\textsc{iv}]~$\lambda1483/\lambda1487$. The UV intercombination diagnostics are available only in a subset of local star-forming galaxies~\cite{Berg:16,Kurt:99,Mingozzi:22} and high-redshift systems~\cite{Welch:24,Topping:24}. At low spectral resolution, [Ar\,\textsc{iv}]~$\lambda4711$ can be blended with He\,\textsc{i}~$\lambda4713$; special care was taken to flag and exclude affected spectra following the approach of ~\cite{MendezDelgado:24FeO}. Uncertainties were propagated through Monte Carlo realizations perturbing the observed line fluxes according to their reported measurement errors. As in the LVM analysis of the Orion Nebula, the electron temperature from [N\,\textsc{ii}]~$\lambda5755/\lambda6584$ was imposed when computing densities from the auroral-to-nebular ratios of [S\,\textsc{ii}] and [O\,\textsc{ii}], ensuring consistency between both datasets~\cite{MendezDelgado:2023a,Rickards:2024}.

For the present analysis, only H\,\textsc{ii} regions, local star-forming galaxies and high-redshift star-forming galaxies were retained; Orion Nebula spectra were excluded to avoid duplication with the LVM dataset. Objects were required to have at least two density diagnostics available. For local H\,\textsc{ii} regions and SFGs, we additionally required a coverage of $\Delta\log_{10} n_\mathcal{M} \geq 2$~dex, ensuring that the density hierarchy can be meaningfully traced within each individual object. For high-redshift SFGs the threshold was relaxed to $\Delta\log_{10} n_\mathcal{M} \geq 0.5$~dex given the limited number of accessible diagnostics at those redshifts. The resulting sample comprises 42 H\,\textsc{ii} regions~\cite{GarciaRojas:04,GarciaRojas:05,GarciaRojas:06,GarciaRojas:07,Esteban:02,Esteban:09,Esteban:13,Esteban:17,Esteban:18,Esteban:20,TorresPeimbert:89,Zurita:12,DominguezGuzman:22,Valerdi:19,PenaGuerrero:12,Peimbert:03}, 58 local star-forming galaxies~\cite{LopezSanchez:07,Esteban:14,Hagele:06,Hagele:08,Kurt:99,Peimbert:12,Guseva:12,Izotov:21b,Berg:16,Fernandez:18,Mingozzi:22} and 3 high-redshift systems~\cite{Welch:24,Topping:24}, together with [O\,\textsc{iii}]~$\lambda4363/\lambda5007$ measurements from~\cite{Arellano:26} for which this ratio is the only density diagnostic accessible. Individual density measurements for each object are provided in a machine-readable table, which will be deposited in a public repository upon acceptance. %Individual density measurements for each object are provided in a machine-readable table at \href{https://drive.google.com/file/d/1WZ3gHWXLkMZFP0rr3B169-GJmx8O1oor/view?usp=sharing}{[temporary link for co-author review; Zenodo DOI to be assigned before submission]}.

\subsection*{Maximum-sensitivity density}

For each density-sensitive line ratio $R$, the maximum-sensitivity density $n_\mathcal{M}$ is defined as the electron density at which the logarithmic sensitivity of the ratio is maximal (equation~\ref{eq:nM}). It was evaluated numerically from \textsc{PyNeb} emissivities at $T_e = 10^4$~K on a grid of $10^4$ points uniformly spaced in $\log n_e$ from $10^0$ to $10^{11}$~cm$^{-3}$, with $|d\log R/d\log n_e|$ computed by finite differences. Peaks in the sensitivity function were identified using the \textsc{scipy} \texttt{find\_peaks} algorithm with a minimum prominence threshold of 0.02. For diagnostics with a unimodal sensitivity function, $n_\mathcal{M}$ corresponds to the unique global peak. In the two-level approximation, $n_\mathcal{M}$ can be shown analytically to equal the geometric mean of the critical densities of the two lines involved, $n_\mathcal{M} = (n_1\,n_2)^{1/2}$, providing a physically transparent interpretation: $n_\mathcal{M}$ marks the density at which neither line is in the low- nor high-density limit, so that the ratio is most sensitive to changes in $n_e$.

For diagnostics with multiple peaks --- including three [Fe\,\textsc{iii}] ratios ($\lambda4658/\lambda4702$, $\lambda4658/\lambda4881$, $\lambda4702/\lambda4881$) and [Fe\,\textsc{ii}]~$\lambda8617/\lambda7155$, whose complex level structures produce bimodal sensitivity curves --- only peaks located above the minimum critical density of the lines involved are considered, and among these $n_\mathcal{M}$ is taken as the peak with the highest sensitivity value. This criterion selects the density regime most physically relevant to the line formation, excluding spurious low-density peaks that arise from atomic level structure rather than genuine density sensitivity. All values are listed in Extended Data Table~\ref{tab:diagnostics}.

\subsection*{Forward model}

The forward model is intentionally minimal: its purpose is not to reproduce the full complexity of a real H\,\textsc{ii} region, but to isolate the physical mechanism responsible for the density hierarchy --- namely, the interaction between a broad unresolved density distribution and the intrinsic response functions of the diagnostics. No ionization stratification, phase structure, spatial correlations or radiation transfer are imposed.

The model constructs a three-dimensional nebula on a $50 \times 50 \times 50$ grid. Electron densities are drawn independently at each grid point from a power-law emission-measure distribution,
\begin{equation}
H_n \propto n_e^{\,\beta}, \quad n_e \in [n_\mathrm{min},\, n_\mathrm{max}],
\label{eq:powerlaw_methods}
\end{equation}
where $H_n\,dn \equiv n_e^2\,dV$ is the contribution of gas in the density interval $[n_e, n_e + dn]$ to the total emission measure, $\beta$ is the power-law slope, and $n_\mathrm{min}$, $n_\mathrm{max}$ are the density bounds. The power law is adopted as the simplest functional form, with the minimum number of parameters, capable of describing a broad continuous density distribution; the slope $\beta$ is left as a free parameter to be constrained by the data. A power law of this kind is produced, for example, by a steady constant-velocity wind, for which the slope is set by how rapidly the flow cross-section diverges: $\beta = -1/2$ for spherical and $\beta = -1$ for cylindrical divergence, with steeper slopes corresponding to flatter, sub-cylindrical geometries~\cite{TenorioTagle:79,Henney:05}. We invoke such flows only as an illustration that physically plausible mechanisms can generate a power-law emission-measure distribution --- in the same sense that turbulence is often invoked to motivate a log-normal PDF --- and not as a claim that the gas is organized into a single ordered flow. Real H\,\textsc{ii} regions are expected to combine many such flows together with their mutual interactions, which largely erase any one-to-one relation between density and distance from the ionization front; the adopted power law therefore carries no implied correlation between $n_e$ and ionization state, consistent with the observation that the density hierarchy persists within a single ionic species. Under this prescription the mass per logarithmic density interval scales as $dM/d\ln n_e \propto H_n \propto n_e^{\beta}$ and therefore decreases with density for $\beta < 0$, so that the densest gas carries little mass and occupies little volume, yet contributes substantially to the forbidden-line luminosity because its intrinsic emissivity is high.
Electron temperatures are drawn from a narrow Gaussian,
\begin{equation}
T_e \sim \mathcal{N}(T_0), \quad T_e \geq 1000\,\mathrm{K},
\label{eq:Te_methods}
\end{equation}
with $T_0 = 10^4$~K and $t^2 = 5 \times 10^{-5}$, producing a model that is essentially isothermal. The upper density is truncated at $\log_{10} n_e = 6$ for definiteness; as shown by the Bayesian analysis below, the recovered relation is insensitive to the precise value of this upper cutoff, since gas well above $n_\mathrm{min}$ contributes negligibly to the emergent luminosity. Such high densities are in any case rare in the diffuse ionized gas considered here, being reached only in extreme compact environments such as photoevaporative protoplanetary disks (proplyds)~\cite{Odell:94,Henney:98,Henney:99}. The temperature field is deliberately kept isothermal to ensure that any density hierarchy emerging from the model is attributable solely to the density structure and the atomic response functions, with no contribution from thermal gradients. 

More sophisticated treatments that couple density and temperature structure self-consistently, such as the photoionization framework of \cite{Marconi:24}, are potentially valuable for exploring the thermal consequences of realistic density distributions and are left for future work.

\begin{figure*}
\centering
\includegraphics[width=0.92\textwidth]{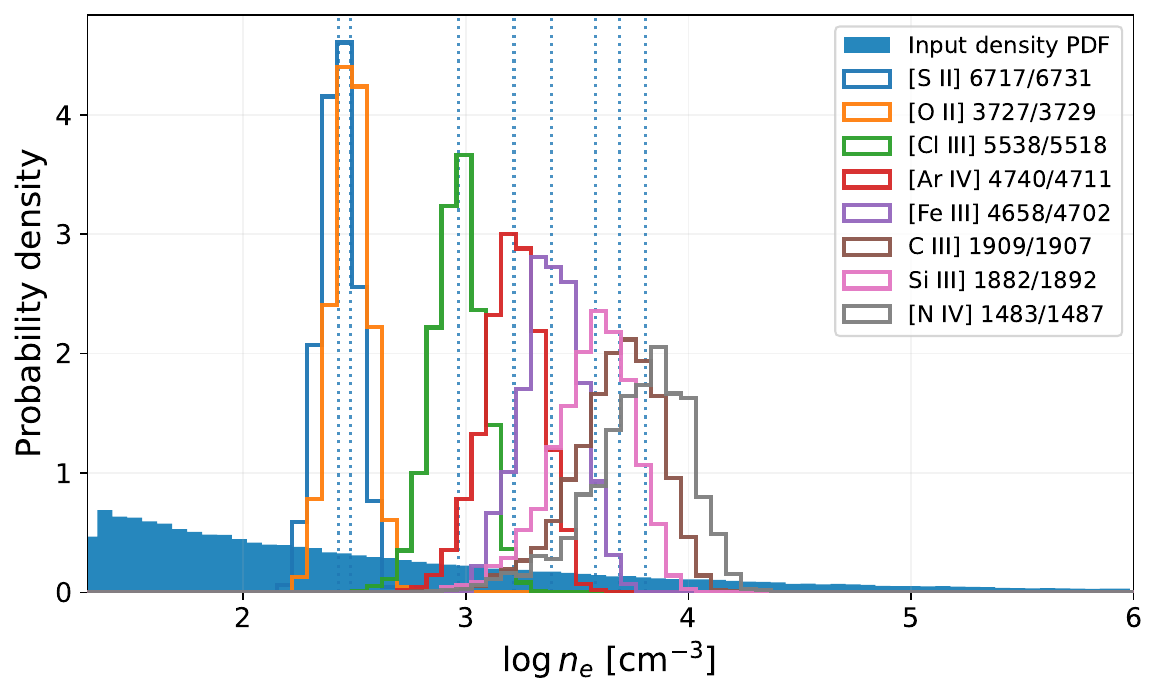}
\caption{\textbf{Distributions of inferred electron densities from the forward model.}
Probability density distributions of $\log_{10} n_e$ recovered by each diagnostic from the forward model with a power-law emission-measure PDF ($\beta = -1.3$, $\log_{10} n_\mathrm{min} = 1.3$, $\log_{10} n_\mathrm{max} = 6.0$), compared with the input distribution (filled blue histogram). Each coloured histogram corresponds to a different diagnostic. The progressive shift of each distribution toward higher densities with increasing $n_\mathcal{M}$ directly visualizes the density-selection bias: every diagnostic recovers a different emissivity-weighted projection of the same underlying density field, and none of them recovers the true volume-weighted mean. Vertical dashed lines mark the median of each distribution.}
\label{fig:extdata_biases}
\end{figure*}

The key insight is that, for a given emission-measure distribution $H(n_e)$, the line-of-sight integrated ratio of any diagnostic can be written as
\begin{equation}
R_\mathrm{obs} = \frac{\int H(n_e)\,\epsilon_\mathrm{num}(n_e, T_0)\,dn_e}{
\int H(n_e)\,\epsilon_\mathrm{den}(n_e, T_0)\,dn_e},
\label{eq:Robs}
\end{equation}
where $H(n_e)$ is the emission-measure PDF defined in equation~(\ref{eq:powerlaw_methods}) and $\epsilon_\lambda$ is the per-ion emissivity returned by \textsc{PyNeb}~\cite{Luridiana:15}. The volume emission coefficient of each line is $j_\lambda \propto n_e\,n_{X^i}\,\epsilon_\lambda$, so its luminosity is $L_\lambda \propto \int n_e\,n_{X^i}\,\epsilon_\lambda\,dV$; for a fixed ionic fraction this is $\propto \int n_e^2\,\epsilon_\lambda\,dV = \int H(n_e)\,\epsilon_\lambda\,dn_e$. The emission-measure weighting $n_e\,n_{X^i}\approx n_e^2$ is therefore already absorbed into $H(n_e)$ through its definition $H_n\,dn \equiv n_e^2\,dV$ in equation~(\ref{eq:powerlaw_methods}): the integration variable is the emission measure rather than the volume, and no separate $n_e^2$ factor or volume element multiplies $\epsilon_\lambda$. For $\beta \approx -1.3$, the distribution $H_n \propto n_e^{-1.3}$ is such that the densest gas contributes to the integrated luminosity roughly as much as the diffuse gas despite occupying little volume. The line ratio $R_\mathrm{obs}$ nevertheless retains density information through the differential dependence of $\epsilon_\mathrm{num}/\epsilon_\mathrm{den}$ on $n_e$, which is the response function $|d\log R/d\log n_e|$ of each diagnostic. The observed ratio $R_\mathrm{obs}$ is therefore not evaluated at any single density but represents a weighted projection of the PDF through the response function of each diagnostic. The resulting distributions of recovered densities are shown in Extended Data Fig.~\ref{fig:extdata_biases}, where the progressive shift of each diagnostic toward higher densities with increasing $n_\mathcal{M}$ is directly visible. For each diagnostic, the numerator and denominator emissivity cubes are summed independently along one spatial axis and their ratio inverted using \textsc{PyNeb} \texttt{getTemDen} at $T_0$, exactly as in observational analyses (Figs.~\ref{fig:model_input} and~\ref{fig:model_sequence}).

The model was validated with a narrow-PDF test: when the density range is restricted to a narrow interval, the underlying density distribution is sufficiently compact that all diagnostics sample essentially the same region of the PDF regardless of their $n_\mathcal{M}$. In this case all diagnostics recover the same median density, and no ordered hierarchy emerges (Extended Data Fig.~\ref{fig:extdata_nobiases}). This confirms that the density hierarchy requires a genuinely broad density field and does not arise from numerical artefacts of the inversion procedure.

\subsection*{Analytic solution for the power-law model}

The 3D forward model isolates the mechanism but is too expensive to embed in a parameter search. In the two-level approximation, however, the line-of-sight integrated ratio of equation~(\ref{eq:Robs}) admits a closed form. The volume emission coefficient of a line with critical density $n_k$ scales as $j_k \propto n_e\,n_{X^i}\,(1+n_e/n_k)^{-1}$, so for a density-sensitive doublet whose two lines have critical densities $n_1$ and $n_2$, the emission-measure weighted ratio becomes
\begin{equation}
R_\mathrm{obs} = R_\mathrm{lo}\,
\frac{\displaystyle\int_{n_\mathrm{min}}^{n_\mathrm{max}}
      \bigl(1+\delta\,n_e/n_\mathcal{M}\bigr)^{-1}\,H_n\,dn_e}
     {\displaystyle\int_{n_\mathrm{min}}^{n_\mathrm{max}}
      \bigl(1+\delta^{-1}\,n_e/n_\mathcal{M}\bigr)^{-1}\,H_n\,dn_e},
\label{eq:twolevel_ratio}
\end{equation}
where $R_\mathrm{lo}$ is the low-density limit of the ratio,
$n_\mathcal{M}=(n_1 n_2)^{1/2}$ is the maximum-sensitivity density (the geometric mean of the two critical densities; see ``Maximum-sensitivity density''), and $\delta=(n_2/n_1)^{1/2}$ is the contrast factor that sets the dynamic range of the diagnostic. For the power-law emission-measure distribution $H_n \propto n_e^{\beta}$ of equation~(\ref{eq:powerlaw_methods}), the normalization cancels and equation~(\ref{eq:twolevel_ratio}) evaluates to
\begin{equation}
R_\mathrm{obs} = R_\mathrm{lo}\,
\frac{n_\mathrm{max}^{\beta+1}\,\mathcal{F}\!\left(\beta,\,\delta\,n_\mathrm{max}/n_\mathcal{M}\right)
      - n_\mathrm{min}^{\beta+1}\,\mathcal{F}\!\left(\beta,\,\delta\,n_\mathrm{min}/n_\mathcal{M}\right)}
     {n_\mathrm{max}^{\beta+1}\,\mathcal{F}\!\left(\beta,\,\delta^{-1}n_\mathrm{max}/n_\mathcal{M}\right)
      - n_\mathrm{min}^{\beta+1}\,\mathcal{F}\!\left(\beta,\,\delta^{-1}n_\mathrm{min}/n_\mathcal{M}\right)},
\label{eq:hypergeom}
\end{equation}
where $\mathcal{F}(\beta,x) \equiv {}_2F_1\!\left(1,\,\beta+1;\,\beta+2;\,-x\right)$ is a Gauss hypergeometric function. We verified that equation~(\ref{eq:hypergeom}) reproduces direct numerical integration of equation~(\ref{eq:twolevel_ratio}) to machine precision. Inverting $R_\mathrm{obs}$ with the standard two-level relation yields the apparent density that each diagnostic would recover from a given power-law distribution, exactly as in the observational analysis, and provides the fast forward evaluation used in the Bayesian fit below.

\subsection*{Bayesian constraints on the density distribution}

\begin{figure*}
\centering
\includegraphics
[width=0.5\linewidth]
{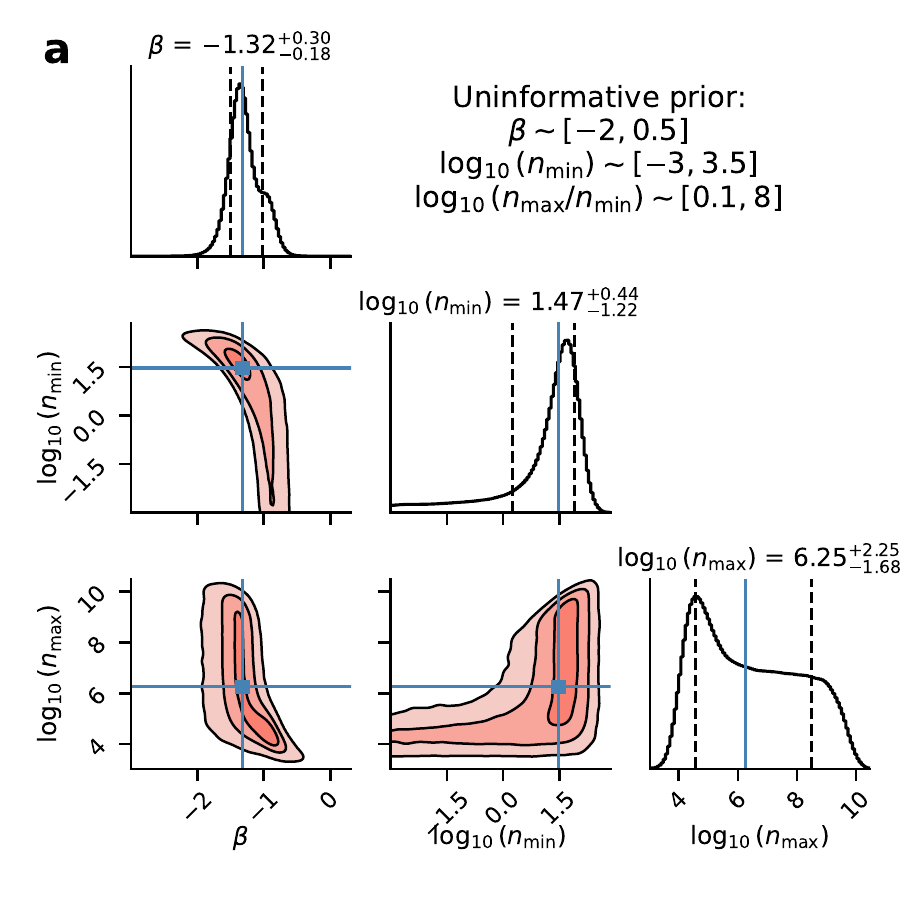}%
\includegraphics
[width=0.5\linewidth]
{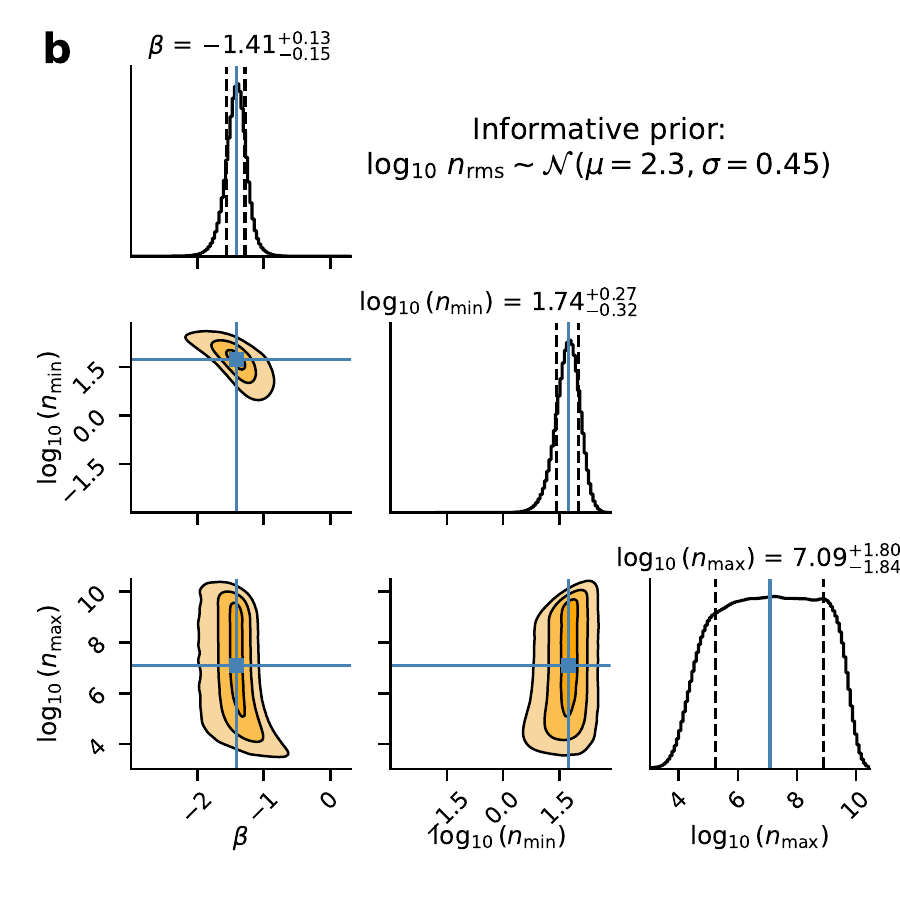}
\caption{\textbf{Posterior distribution of the power-law density-model parameters.}
Marginalized one- and two-dimensional posterior distributions for the three parameters of the power-law emission-measure model (equation~\ref{eq:powerlaw_methods}), obtained by fitting the nine independent Orion diagnostics with the \textsc{emcee} sampler (see Methods): the slope $\beta$, the lower density bound $\log_{10} n_\mathrm{min}$, and upper bound $\log_{10} n_\mathrm{max}$. Dashed lines and titles give the median and the 16th--84th percentile range of each parameter; blue lines mark the median values. 
(\textbf{a})~Results from employing broad, uninformative priors 
for \(\beta\), \(\log_{10} n_\mathrm{min}\), and \(\log_{10}(n_\mathrm{max}/n_\mathrm{min})\).
The slope and lower density bound are well constrained, whereas the upper density is constrained only from below and its upper tail runs into the hard prior cutoff on \(\log_{10}(n_\mathrm{max}/n_\mathrm{min})\), reflecting the insensitivity of the data to how far the dense tail extends.
(\textbf{b})~Results from employing an informative prior 
on the RMS electron density, 
\(\log_{10} n_\text{rms} \sim \mathcal{N} (2.3, 0.45^2)\),
based on the Str\"omgren condition for the Orion Nebula (see Methods).
This effectively removes the low \(n_{\mathrm{min}}\) tail at \(\beta \approx -1\),
since it corresponds to \(\log_{10} n_{\mathrm{rms}} \ll 2\), which is
incompatible with the observed radius of the Orion Nebula.
As a result, the posterior distributions of \(\beta\) and \(\log_{10} n_{\mathrm{min}}\)
are slightly shifted and tightened by a factor of about two,
while the upper density bound has once more only a lower limit that is physically meaningful,
\(\log_{10} n_{\mathrm{max}} > 5.2\), with the high end still unconstrained by the data.}
\label{fig:extdata_corner}
\end{figure*}

To quantify how well the observed density hierarchy constrains the underlying density field, we fitted the power-law emission-measure model of equation~(\ref{eq:powerlaw_methods}) directly to the Orion sequence. The fit was performed in log--log space on the nine algebraically independent diagnostics, each assigned an uncertainty of $0.25$~dex in $\log_{10} n_{e,\mathrm{obs}}$. For a given set of parameters $(\beta,\,\log_{10} n_\mathrm{min},\,\log_{10}(n_\mathrm{max}/n_\mathrm{min}))$, the apparent density of each diagnostic was predicted from the closed-form solution of equation~(\ref{eq:hypergeom}) and inverted exactly as in the observational analysis. Rather than assigning each diagnostic its exact contrast factor $\delta$, we adopted the empirical scaling $\delta = 2\,(n_\mathcal{M}/10^{3}\,\mathrm{cm}^{-3})^{1/2}$, calibrated to the range spanned by the diagnostics from [S\,\textsc{ii}]~$\lambda6717/\lambda6731$ ($\delta \approx 2$, $\log_{10} n_\mathcal{M} \approx 3$) to C\,\textsc{iii}]~$\lambda1907/\lambda1909$ ($\delta \approx 200$, $\log_{10} n_\mathcal{M} \approx 7$). This captures the systematic increase of diagnostic dynamic range with $n_\mathcal{M}$ while keeping the forward model fast; the recovered density hierarchy is set primarily by $n_\mathcal{M}$ rather than by the precise value of $\delta$. We first adopted broad uniform priors,  $\beta \in [-3,\,0.5]$, $\log_{10} n_\mathrm{min} \in [-3,\,3.5]$ and $\log_{10}(n_\mathrm{max}/n_\mathrm{min}) \in [0.1,\,12]$, parameterising the upper bound through the width $n_\mathrm{max}/n_\mathrm{min}$ in order to enforce $n_\mathrm{max} > n_\mathrm{min}$, although all results are reported in terms of \(\log_{10} n_\text{max}\). Posterior distributions were sampled with the \textsc{emcee} affine-invariant ensemble sampler~\cite{ForemanMackey:13} as implemented in \textsc{lmfit}~\cite{Newville:14}, using $10^4$ steps per walker, a burn-in of $10^3$ steps and a thinning factor of 1. The resulting posterior is shown in Extended Data Fig.~\ref{fig:extdata_corner}a.

The slope and lower density bound are well constrained, $\beta = -1.32\,^{+0.30}_{-0.18}$ and $\log_{10} n_\mathrm{min} = 1.47\,^{+0.44}_{-1.22}$, in agreement with the value $\beta \approx -1.3$ adopted in the forward model. The upper density bound, by contrast, is bounded only from below, $\log_{10} n_\mathrm{max} > 4.6$ (68\% credibility; $> 3.9$ at 95\%), with its upper tail running into the prior ceiling on \(\log_{10}(n_\mathrm{max}/n_\mathrm{min})\). This is the expected behaviour: once the distribution extends a few dex above $n_\mathrm{min}$, gas at still higher density contributes negligibly to the emergent luminosity, so the data cannot determine how far the tail extends. 

Close inspection of Fig.~\ref{fig:extdata_corner}a shows that the posterior distribution of the power law slope is bimodal, 
with a primary peak at \(\beta \approx -1.4\) and a secondary peak at \(\beta \approx -1\). 
However, the latter is associated with very small values of the lower density bound: \(\log_{10} n_\text{min} < 0\), 
which may not be realistic.
In order to test this, we introduced an additional constraint into our priors based on the root-mean-square ionized density, \(n_\text{rms}\), of the nebula. 
Under the assumption of static photoionization equilibrium, one finds the Strömgren condition: \(n_\text{rms} = \left[ 3 Q_\text{eff} / 4 \pi \alpha_\text{B} R_\text{IF}^3 \right]^{1/2}\),
where \(Q_\text{eff}\) is the effective Lyman continuum luminosity of the ionizing stars (discounting the fraction that escapes or is absorbed by dust grains),
\(\alpha_\text{B}\) is the Case~B recombination coefficient,
and \(R_\text{IF}\) is the radius of the ionization front.
Taking the spread of plausible values for these quantities in the Orion Nebula \cite{Simon-Diaz:06, ODell:09, ODell:10},
we find \(\log_{10} n_\text{rms} \simeq 2.30 \pm 0.45\).
We therefore repeat the Bayesian analysis with an informative prior
given by a normal distribution of \(\log_{10} n_\text{rms} \sim \mathcal{N}(\mu = 2.30, \sigma = 0.45)\),
with results shown in Fig.~\ref{fig:extdata_corner}b.

It can be seen that the additional restriction completely eliminates the secondary \(\beta \approx -1\) peak,
leading to tighter constraints on the parameters:
\(\beta = -1.41^{+0.13}_{-0.15}\), \(\log_{10} n_\text{min} = 1.74^{+0.27}_{-0.32}\),
although the upper density bound remains a lower limit:
\(\log_{10} n_\text{max} > 5.2\) (68\% credibility; \(> 4.2\) at 95\%).
Integrating the posterior emission-measure distribution, one finds that $(10^{+5}_{-4})\%$  of the recombination-line emission originates from gas with $\log_{10} n_e \gtrsim 4$,
and $(3^{+3}_{-2})\%$ 
from gas with $\log_{10} n_e \gtrsim 5$.
The bulk gas density is therefore well
established, whereas the densest component is bounded only from below --- it must reach $\log_{10} n_e \gtrsim 5$, but may extend to arbitrarily high density while contributing a vanishing fraction of the light.

\subsection*{The absence of a density hierarchy in kinematically isolated gas}

\begin{figure*}
\centering
\includegraphics[width=0.92\textwidth]{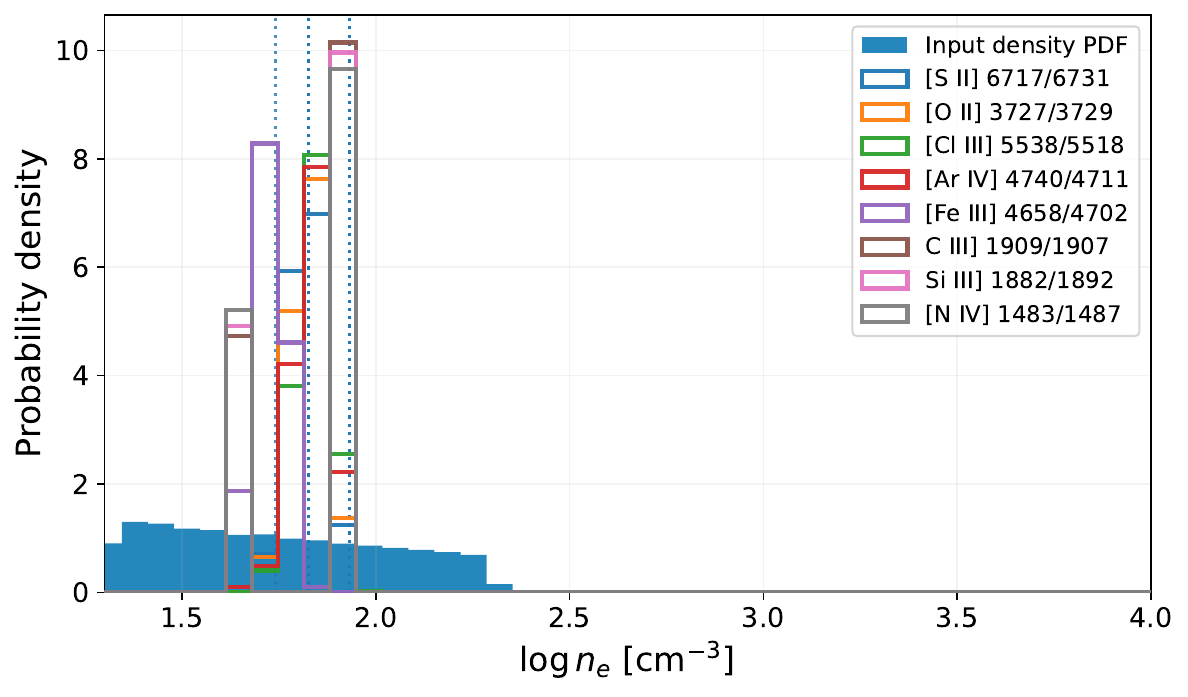}
\caption{\textbf{Narrow-PDF model validation.}
Probability density distributions of $\log_{10} n_e$ recovered by each diagnostic from a forward model identical to that of Extended Data Fig.~\ref{fig:extdata_biases} but with a narrow power-law emission-measure distribution spanning only one order of magnitude in density ($\beta = -1.3$, $\log_{10} n_\mathrm{min} = 1.3$, $\log_{10} n_\mathrm{max} = 2.3$) instead of the broad distribution spanning nearly five orders of magnitude used in the main analysis. All diagnostics recover essentially the same median density and no ordered hierarchy emerges, confirming that the density hierarchy requires a non-negligible high-density contribution to the line luminosity and does not arise from numerical artefacts of the ratio-inversion procedure.}
\label{fig:extdata_nobiases}
\end{figure*}

The forward model predicts that the density hierarchy should disappear when the gas sampled along the line of sight has a narrow density distribution (Extended Data Fig.~\ref{fig:extdata_nobiases}). This limiting case is realized in the photoionized Herbig-Haro objects HH\,202-S and HH\,204 in the Orion Nebula, analysed at high spectral resolution with VLT/UVES echelle spectroscopy~\cite{MesaDelgado:09,MendezDelgado:21}. Each bow shock moves supersonically relative to the ambient gas, Doppler-shifting its emission away from the systemic nebular velocity; the high resolution then separates the dense, compact shocked gas of the HH object from the foreground and background nebular emission along the same sightline. Crucially, the proper motions of both objects show that they propagate close to the plane of the sky --- the velocity vector lies within $\sim$$16^\circ$ of the plane for HH\,204 and $\sim$$48^\circ$ for HH\,202-S~\cite{ODell:08,MendezDelgado:21} --- so that the line of sight intersects only the thin, dense shocked shell rather than the full length of the flow, limiting the depth over which density varies and ensuring that the isolated emission samples a genuinely narrow density range. The isolated component therefore samples a single narrow range of densities rather than the broad distribution integrated through the full nebula.

\begin{figure*}
\centering
\includegraphics[width=0.92\textwidth]{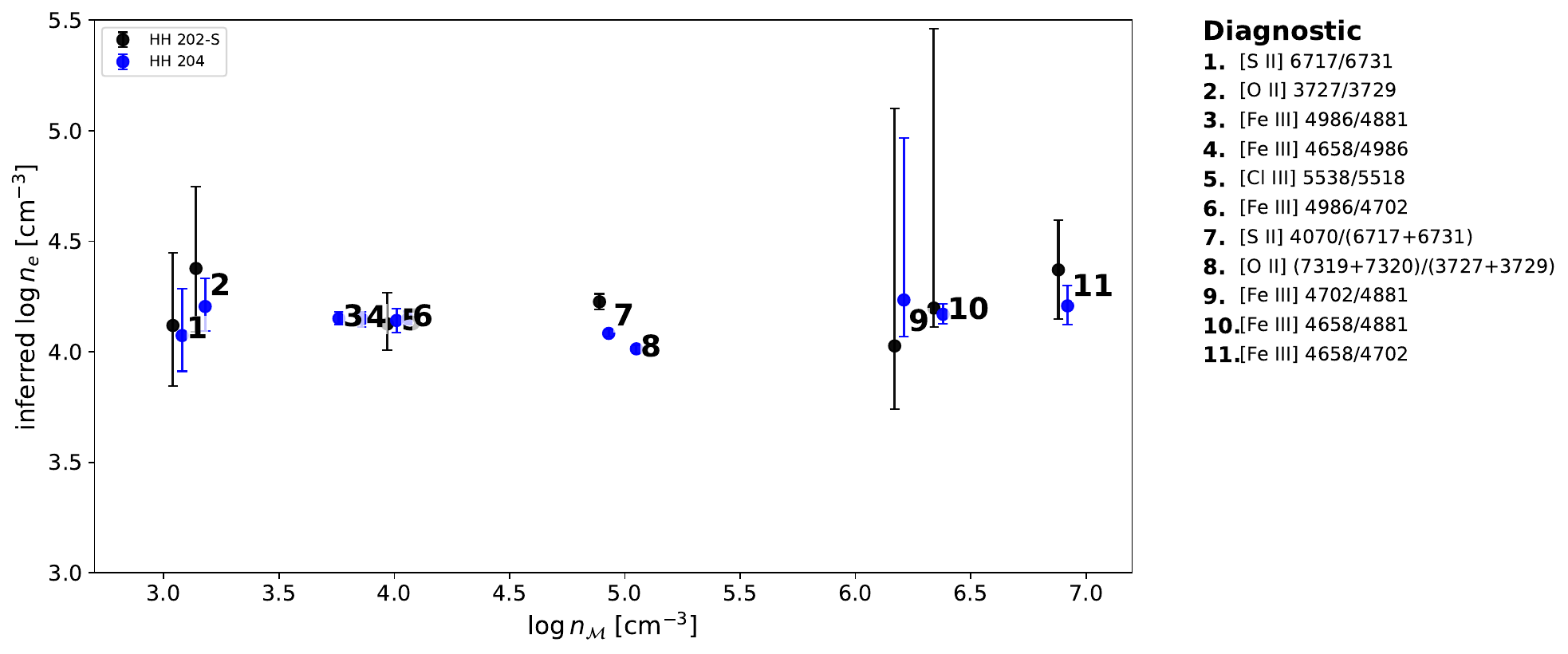}
\caption{\textbf{Observational counterpart of the narrow-PDF test: the hierarchy vanishes in a kinematically isolated dense component.}
Inferred electron density as a function of the maximum-sensitivity density $n_\mathcal{M}$ for the photoionized Herbig-Haro objects HH\,202-S (black) and HH\,204 (blue) in the Orion Nebula, whose high-velocity emission is Doppler-separated from the surrounding nebula through high-resolution echelle spectroscopy~\cite{MesaDelgado:09,MendezDelgado:21}. Because each bow shock moves supersonically relative to the ambient gas, its dense, compact emitting region ($\log_{10} n_e \approx 4.1$) is cleanly isolated along the line of sight. In contrast to the integrated Orion sequence (Fig.~\ref{fig:main_sequence}), all eleven diagnostics --- spanning nearly four orders of magnitude in $n_\mathcal{M}$, from [S\,\textsc{ii}] and [O\,\textsc{ii}] to high-$n_\mathcal{M}$ [Fe\,\textsc{iii}] ratios --- return the same density within the uncertainties, with no trend against $n_\mathcal{M}$. This is the observational realization of the narrow-PDF test of Extended Data Fig.~\ref{fig:extdata_nobiases}: when the gas sampled along the line of sight has a genuinely narrow density distribution, the density-selection bias disappears. Error bars span the 16th--84th percentile range; diagnostic numbering follows Extended Data Table~\ref{tab:diagnostics}.}
\label{fig:extdata_hh}
\end{figure*}

For this isolated component the diagnostic hierarchy vanishes: the same ratios that order by $n_\mathcal{M}$ in the integrated nebula instead converge on a common density of $\log_{10} n_e \approx 4.1$, from [S\,\textsc{ii}] and [O\,\textsc{ii}] to the high-$n_\mathcal{M}$ [Fe\,\textsc{iii}] ratios, with no trend across nearly four orders of magnitude in $n_\mathcal{M}$ (Extended Data Fig.~\ref{fig:extdata_hh}). Conversely, degrading the spectral resolution to blend these dense knots back with the foreground nebula restores the discrepancy, the low-$n_\mathcal{M}$ ratios collapsing to the bulk density while the high-$n_\mathcal{M}$ ratios remain high~\cite{MendezDelgado:21}. This is the observational demonstration of Extended Data Fig.~\ref{fig:extdata_nobiases}: the hierarchy is governed by whether the line-of-sight density distribution is narrow or broad, and not by atomic data, which are unchanged between the two cases, nor by ionization stratification, which cannot vanish for a single component emitting across these ionization stages.

\subsection*{Comparison with the cold-ISM density distribution}

Power-law density distributions are well documented for the cold, molecular ISM, both observationally~\cite{Schneider:15} and in simulations of gravoturbulent, star-forming gas~\cite{FederrathKlessen:13,KlessenGlover:16}, typically as the high-density power-law tail of an otherwise log-normal distribution~\cite{Chen:18}. These studies, however, generally characterize the volume-weighted distribution of the column density $N$, which is not the same quantity as the emission-measure-weighted ($n_e^2\,dV$), linear distribution of the volume density $n_e$ constrained here~\cite{Brunt:10}; the two are connected only through assumptions about geometry and line-of-sight structure, so their slopes cannot be compared directly.

A convenient common ground is the equivalent radial density profile $\rho \propto r^{-\alpha}$. For such a profile in $D$ dimensions, the volume per unit density interval is $dV/dn_e \propto n_e^{-D/\alpha - 1}$, while our emission-measure distribution $H_n \propto n_e^{\beta}$ implies $dV/dn_e \propto n_e^{\beta - 2}$ (since $H_n\,dn_e \equiv n_e^2\,dV$). Equating the two exponents gives
\begin{equation}
\alpha = \frac{D}{1 - \beta}.
\label{eq:alpha_beta}
\end{equation}
Equivalently, the volume-density distribution $dV/dn_e \propto n_e^{\beta-2}$ has the power-law slope expected for a radial profile $\rho \propto r^{-\alpha}$, the standard correspondence used in the gravoturbulent framework~\cite{FederrathKlessen:13}. For $\beta \approx -1.3$ this yields $\alpha \approx 1.3$ for spherical ($D=3$) and $\alpha \approx 0.87$ for cylindrical ($D=2$) divergence. Compared like-for-like (spherical to spherical), our $\alpha \approx 1.3$ is somewhat shallower than the $\alpha \approx 1.5$--$2$ inferred for molecular clouds~\cite{Schneider:15}.

We do not regard this offset as problematic, because the ionized density field is expected to resemble, but not to reproduce, that of the neutral gas from which it forms. Only gas flash-ionized by a fast R-type front retains the pre-existing neutral density structure, and this configuration is transient, lasting at most a recombination time ($\lesssim 10^3$~yr), so it cannot be globally representative. The prevailing D-type fronts impose only a loose, non-monotonic relation between neutral and ionized density: photoevaporation preferentially ionizes and disperses the smallest, densest condensations, and the pressure imbalance following ionization further redistributes the gas. Under simple pressure balance between the two phases the power-law slope is preserved but shifted to lower densities. A quantitative mapping between the neutral and ionized density distributions, which requires self-consistent ionization-front modelling, is left for future work.

\subsection*{Atomic data}

All emissivity calculations used the atomic dataset described in ~\cite{MendezDelgado:2023a}, loaded via the \textsc{PyNeb} atomic data infrastructure.

%% ============================================================
%% ACKNOWLEDGEMENTS & DECLARATIONS
%% ============================================================

\bmhead{Acknowledgements}
The authors thank Michael G.\ Richer for his valuable comments and feedback on the manuscript. J.E.M.-D., C.M., W.J.H., S.F.S., H.J.I.-M., R.O.-D., A.Z.L.-A., A.W., R.d.J.Z. and L.C.C.-C. gratefully acknowledge support from the Secretar\'ia de Ciencia, Humanidades, Tecnolog\'ia e Innovaci\'on (SECIHTI) project CBF-2025-I-2048, ``Resolviendo la F\'isica Interna de las Galaxias: De las Escalas Locales a la Estructura Global con el SDSS-V Local Volume Mapper''. J.E.M.-D., C.M., W.J.H., H.J.I.-M., C.E., J.G.-R., A.Z.L.-A., R.d.J.Z. and L.C.C.-C. gratefully acknowledge support from the UNAM/DGAPA/PAPIIT IA103326 project ``DESIRED (DEep Spectra of ionised Regions Database): de las emisiones m\'as sutiles a la f\'isica fundamental del universo''.
W.J.H. gratefully acknowledges financial support provided by
Direcci\'on General de Asuntos del Personal Acad\'emico, Universidad Nacional Aut\'onoma de M\'exico,
through grants
``Programa de Apoyo a Proyectos de Investigaci\'on e Innovaci\'on Tecnol\'ogica IN111124, IN117326''.
C.E. and J.G.-R. acknowledge support from the Agencia Estatal de Investigaci\'on of the Ministerio de Ciencia, Innovaci\'on y Universidades (AEI-MCIU) under grant `The internal structure of ionised nebulae and its effects in the determination of the chemical composition of the interstellar medium and the Universe' with reference PID2023-151648NB-I00 (DOI:10.13039/5011000110339).
S.C.O.G. acknowledges financial support from the ERC via Synergy Grant ``ECOGAL'' (project ID 855130) and from the German Excellence Strategy via the Heidelberg Cluster ``STRUCTURES'' (EXC 2181 - 390900948).
K.K., F.-H.L., N.S. and E.E. acknowledge funding from the European Research Council's Starting Grant ERC StG-101077573 (`ISM-METALS'). O.V.E. acknowledges funding from the Deutsche Forschungsgemeinschaft (DFG, German Research Foundation) -- project-ID 541068876. I.A.Z. acknowledges funding from the Deutsche Forschungsgemeinschaft (DFG, German Research Foundation) -- project-ID 550945879. Funding for the Sloan Digital Sky Survey V has been provided by the Alfred P. Sloan Foundation, the Heising-Simons Foundation, the National Science Foundation, and the Participating Institutions. SDSS acknowledges support and resources from the Center for High-Performance Computing at the University of Utah. SDSS telescopes are located at Apache Point Observatory, funded by the Astrophysical Research Consortium and operated by New Mexico State University, and at Las Campanas Observatory, operated by the Carnegie Institution for Science. The SDSS web site is \url{https://www.sdss.org}.
SDSS is managed by the Astrophysical Research Consortium for the Participating Institutions of the SDSS Collaboration, including the Carnegie Institution for Science, Chilean National Time Allocation Committee (CNTAC) ratified researchers, Caltech, the Gotham Participation Group, Harvard University, Heidelberg University, The Flatiron Institute, The Johns Hopkins University, L'Ecole polytechnique f\'{e}d\'{e}rale de Lausanne (EPFL), Leibniz-Institut f\"{u}r Astrophysik Potsdam (AIP), Max-Planck-Institut f\"{u}r Astronomie (MPIA Heidelberg), Max-Planck-Institut f\"{u}r Extraterrestrische Physik (MPE), Nanjing University, National Astronomical Observatories of China (NAOC), New Mexico State University, The Ohio State University, Pennsylvania State University, Smithsonian Astrophysical Observatory, Space Telescope Science Institute (STScI), the Stellar Astrophysics Participation Group, Universidad Nacional Aut\'{o}noma de M\'{e}xico, University of Arizona, University of Colorado Boulder, University of Illinois at Urbana-Champaign, University of Toronto, University of Utah, University of Virginia, Yale University, and Yunnan University.

\section*{Declarations}

\textbf{Author contributions.} J.E.M.-D. conceived the study, introduced the maximum-sensitivity density parameter $n_\mathcal{M}$, identified the universal density hierarchy as an intrinsic bias of forbidden-line spectroscopy, processed all LVM observations of the Orion Nebula, derived the electron density maps and physical conditions, performed the DESIRED analysis, and wrote the manuscript. C.M. conceived, designed and implemented the three-dimensional forward model, including the mathematical framework connecting the power-law emission-measure distribution to the emissivity-weighted diagnostic projections. W.J.H. developed the analytical formalism relating the emission-measure density distribution $H_n$ to the integrated diagnostic ratios, the physical scenarios that can give rise to a power-law $H_n$, and performed the Bayesian (MCMC) fit constraining the density-distribution parameters. The remaining authors contributed to the acquisition of the data, to the development and operation of the SDSS-V and LVM survey infrastructure and data systems, discussed the results, and provided critical input on the interpretation of the data and the manuscript

\textbf{Competing interests.} The authors declare no competing interests.

\textbf{Data availability.} LVM data products are publicly available through SDSS-V data releases at \url{https://www.sdss.org}. Individual density measurements for all DESIRED objects are provided in a machine-readable table that will be made publicly available in a data repository upon acceptance. The underlying data are available from the corresponding author upon reasonable request. %Individual density measurements for all DESIRED objects are provided in a machine-readable table at \href{https://drive.google.com/file/d/1WZ3gHWXLkMZFP0rr3B169-GJmx8O1oor/view?usp=sharing}{[temporary link for co-author review; Zenodo DOI to be assigned before submission]}. 
Source data for all figures are available from the corresponding author upon reasonable request.

\textbf{Code availability.} Physical conditions were derived using \textsc{PyNeb} (\url{https://github.com/Morisset/PyNeb\_devel}~\cite{Luridiana:15}) and its machine-learning module \textsc{ai4neb}.

\section*{Extended Data}

\begin{table*}
\caption{\textbf{Forbidden-line density diagnostics used in this work.}
For each diagnostic the table lists: the ionic species and line ratio; the ionization potential (IP) of the parent ion in eV; and the maximum-sensitivity density $\log_{10} n_\mathcal{M}$ [cm$^{-3}$] computed from \textsc{PyNeb} emissivities at $T_e = 10^4$~K. For diagnostics with multiple peaks in the sensitivity function, $n_\mathcal{M}$ corresponds to the peak of highest sensitivity located above the minimum critical density of the lines involved (see Methods). Diagnostics marked $^\ddagger$ are available in the LVM data but not as individual ratios in the DESIRED database~\cite{MendezDelgado:23b}, which instead provides the blended sums $(\lambda\lambda7319{+}7320{+}7330{+}7331)/(\lambda3727{+}\lambda3729)$ and $(\lambda4070{+}\lambda4075)/(\lambda6717{+}\lambda6731)$.
}
\label{tab:diagnostics}
\begin{tabular}{llcc}
\hline
Ion & Ratio & IP [eV] & $\log_{10} n_\mathcal{M}$ [cm$^{-3}$] \\
\hline
{[S\,\textsc{ii}]}               & $\lambda6717/\lambda6731$                                               & 10.4 & 3.06 \\
{[O\,\textsc{ii}]}               & $\lambda3727/\lambda3729$                                               & 13.6 & 3.16 \\
{[Fe\,\textsc{iii}]}             & $\lambda4986/\lambda4881$                                               & 16.2 & 3.74 \\
{[Fe\,\textsc{iii}]}             & $\lambda4658/\lambda4986$                                               & 16.2 & 3.84 \\
{[Cl\,\textsc{iii}]}             & $\lambda5538/\lambda5518$                                               & 23.8 & 3.99 \\
{[Fe\,\textsc{iii}]}             & $\lambda4986/\lambda4702$                                               & 16.2 & 4.04 \\
{[Ar\,\textsc{iv}]}              & $\lambda4740/\lambda4711$                                               & 40.7 & 4.36 \\
{[S\,\textsc{ii}]}$^\ddagger$    & $\lambda4070/\lambda6717$                                               & 10.4 & 4.64 \\
{[O\,\textsc{ii}]}$^\ddagger$    & $\lambda7320/\lambda3729$                                               & 13.6 & 4.69 \\
{[S\,\textsc{ii}]}$^\ddagger$    & $\lambda4070/(\lambda6717+\lambda6731)$                                 & 10.4 & 4.85 \\
{[S\,\textsc{ii}]}               & $(\lambda4070{+}\lambda4075)/(\lambda6717{+}\lambda6731)$               & 10.4 & 4.91 \\
{[O\,\textsc{ii}]}$^\ddagger$    & $\lambda7320/(\lambda3727+\lambda3729)$                                 & 13.6 & 5.03 \\
{[O\,\textsc{ii}]}               & $(\lambda\lambda7319{+}7320{+}7330{+}7331)/(\lambda3727{+}\lambda3729)$ & 13.6 & 5.03 \\
{[O\,\textsc{ii}]}$^\ddagger$    & $\lambda7320/\lambda3727$                                               & 13.6 & 5.10 \\
{[Fe\,\textsc{iii}]}             & $\lambda4702/\lambda4881$                                               & 16.2 & 6.19 \\
{[Fe\,\textsc{ii}]}              & $\lambda8617/\lambda7155$                                               &  7.9 & 6.31 \\
{[Fe\,\textsc{iii}]}             & $\lambda4658/\lambda4881$                                               & 16.2 & 6.36 \\
{[O\,\textsc{iii}]}              & $\lambda4363/\lambda5007$                                               & 35.1 & 6.42 \\
C\,\textsc{iii}]                 & $\lambda1909/\lambda1907$                                               & 47.9 & 6.70 \\
{[Fe\,\textsc{iii}]}             & $\lambda4658/\lambda4702$                                               & 16.2 & 6.90 \\
{[N\,\textsc{iv}]}               & $\lambda1483/\lambda1487$                                               & 47.4 & 7.14 \\
Si\,\textsc{iii}]                & $\lambda1882/\lambda1892$                                               & 33.5 & 7.40 \\
\hline
\end{tabular}
\end{table*}

\bibliography{sn-bibliography}

%% ============================================================
%% FIGURE LEGENDS
%% ============================================================

\clearpage

\end{document}